# Training Data Governance for Brain Foundation Models

Margot Hanley[1]   Jiunn-Tyng Yeh[1]   Ryan Rodriguez[1]

Jack Pilkington[1]   Nita Farahany[2]

[1] Science and Society, Duke University

[2] School of Law, Duke University

## Abstract

Brain foundation models bring the foundation model paradigm to the field of neuroscience. Like language and image foundation models, they are general-purpose AI systems pretrained on large-scale datasets that adapt readily to downstream tasks. Unlike text-and-image based models, however, they train on brain data: large-datasets of EEG, fMRI, and other neural data types historically collected within tightly governed clinical and research settings.

This paper contends that training foundation models on neural data opens new normative territory. Neural data carry stronger expectations of, and claims to, protection than text or images, given their body-derived nature and historical governance within clinical and research settings. Yet the foundation model paradigm subjects them to practices of large-scale repurposing, cross-context stitching, and open-ended downstream application. Furthermore, these practices are now accessible to a much broader range of actors, including commercial developers, against a backdrop of fragmented and unclear governance.

To map this territory, we first describe brain foundation models' technical foundations and training-data ecosystem. We then draw on AI ethics, neuroethics, and bioethics to organize concerns across privacy, consent, bias, benefit sharing, and governance. For each, we propose both agenda-setting questions and baseline safeguards as the field matures.

INTRODUCTION

Over the past few years, the foundation model paradigm has reshaped the field of artificial intelligence (AI). First introduced to the public through OpenAI's release of its chat application ChatGPT,[1] foundation models mark a departure from earlier AI systems. Rather than being built for a single purpose, AI foundation models are defined by their ability to learn general underlying patterns in data and then can be adapted to many downstream applications without extensive retraining,[2] as illustrated by ChatGPT's ability to draft a love poem, edit a legal document, and write computer code. This same foundation paradigm is now being brought to the field of neuroscience in the form of *brain foundation models.*

Brain foundation models are built not on text, as are large language models (LLMs), nor primarily on images, as are diffusion models, but on vast, diverse datasets of brain data[3], such as electroencephalography (EEG) recordings or functional magnetic resonance imaging (fMRI) scans from research studies or routine clinical care.[4] Through training, they build up a rich, statistical mapping, or representation, of typical brain patterns and then use it as a foundation that can be rapidly adapted to a range of tasks, from simulating the trajectory of a person's disease to decoding intended movements or speech. Furthermore, these models are not purely theoretical: while technical advances continue in academic and research labs,[5] neurotechnology and neuroscience companies are already piloting these models in high-stakes clinical contexts and integrating them into next-generation devices. For example, Pyramidal, a Silicon Valley AI start-up, is partnering with the Cleveland Clinic to pilot a model that assists neurologists, in real time, in intensive care units;[6] and a neurotech company Muse has begun to embed and deploy models in their EEG-based headbands for a range of uses, including sleep.[7]

---

[1] Strictly speaking, ChatGPT is a downstream application built on top of OpenAI's GPT-series large language models (LLMs). Those underlying LLMs, rather than the ChatGPT product, exemplify the foundation model paradigm.

[2] R. Bommasani et al., *On the Opportunities and Risks of Foundation Models* (Stanford Ctr. for Research on Foundation Models, 2021), https://crfm.stanford.edu/report.html.

[3] The term "brain data" we used interchangeably with "neural data" throughout the paper refers to data obtained from neurological recording or imaging of both the peripheral and central nervous systems. The common sources of the brain data include but do not limited to electromyography (EMG) electroencephalography (EEG), functional magnetic resonance imaging (fMRI), magnetoencephalography (MEG), functional near-infrared spectroscopy (fNIRS), and brain sonography.

[4] X. Zhou et al., *Brain Foundation Models: A Survey on Advancements in Neural Signal Processing and Brain Discovery*, arXiv (2025), https://arxiv.org/abs/2503.00580.

[5] J. Wu et al., *AdaBrain-Bench: Benchmarking Brain Foundation Models for Brain-Computer Interface Applications*, arXiv (2025), https://arxiv.org/abs/2507.09882 ; W. Xiong et al., *EEG-FM-Bench: A Comprehensive Benchmark for the Systematic Evaluation of EEG Foundation Models*, arXiv (2025), https://arxiv.org/abs/2508.17742.

This paper will look at the ethical, social, and legal issues raised by brain foundation models,[8] with a focus on the data used to train them. While downstream concerns around the *use* of these models, such as context-specific deployment harms, also warrant attention, we concentrate on the sourcing, selection, and governance of training data used to train the models.

Current brain foundation models are trained on large-scale datasets of brain data drawn from two broad sources: (1) public clinical and research datasets and (2) proprietary data streams generated by wearable and invasive neurotechnology devices. Datasets collected decades ago for clinical research are being repurposed for open-ended applications. Developers often stitch these sources into a training dataset, developing a single model on medical datasets collected and governed within tightly bounded clinical and research environments alongside new commercial ones. Such shifts in scale, source, and use constitute a novel training environment for brain data, and this environment is the focus of our inquiry.

Brain foundation models represent a novel intersection of the foundation model paradigm and brain data as a training substrate. Given the limited but rapidly growing technical work on brain foundation models themselves, this paper maps the ethical, social, and legal questions that may emerge at the training data layer by drawing on: (1) what is currently known about how these models are being built and trained, (2) concerns identified in foundation models broadly, including about data scraping, copyright, and bias in text- and image-trained foundation models, and (3) longstanding questions in neuroethics and bioethics about mental privacy and the collection and sharing of sensitive medical and research data.[9]

---

[6] Devin Coldewey, *Piramidal's Foundation Model for Brain Waves Could Supercharge EEGs*, TECHCRUNCH (Aug. 23, 2024), https://techcrunch.com/2024/08/23/piramidals-foundation-model-for-brainwaves-could-supercharge-eegs/. ; Emily Mullin, *An AI Model for the Brain Is Coming to the ICU*, WIRED (Aug. 11, 2025), https://www.wired.com/story/an-ai-model-for-the-brain-is-coming-to-the-icu-cleveland-clinic-piramidal/.
[7] *Muse's AI Platform Delivers Clinical-Grade Sleep and Brain Health Intelligence*, BUSINESS WIRE (June 3, 2025), https://www.businesswire.com/news/home/20250603051086/en/Muses-AI-Platform-Delivers-Clinical-Grade-Sleep-and-Brain-Health-Intelligence.
[8] For brevity, we refer to these collectively as "ethical issues" except when distinguishing legal doctrine from broader normative concerns is important.
[9] In this paper, we focus specifically on neural data, however there are a range of other data modalities such as genomic data, that scholars have pointed to as having similarly distinct benefits and risks. Work has focused on

The paper proceeds in two parts. Part 1 describes how brain foundation models function and can be applied, as well as the diverse datasets that underpin them, from clinical archives to emerging commercial streams. Part 2 builds from that technical grounding to organize key ethical, social, and legal considerations spanning privacy, consent, bias, benefit sharing, and governance. We highlight concerns that may be familiar from adjacent work in AI and biomedical ethics, but are amplified in this setting, alongside concerns that this intersection makes newly possible. In doing so, the paper questions the appropriate use of frameworks that have begun to structure LLM training data, suggesting, for example, a shift away from ownership and copyright towards stewardship, dignity, and collective governance. Our aim is not to resolve these issues here, but to surface and articulate them. Finally, we propose baseline safeguards that can serve as interim measures as the field matures.

Different readers will be drawn to different parts of the paper. Developers of brain foundation models and neurotechnologies will benefit from a map of ethical concerns, directly relevant to their upstream training data choices. AI and AI ethics researchers may see familiar foundation-model concerns recast in the high-stakes context of brain data; neuroethicists will encounter a new ethical terrain in how brain datasets are being repurposed into large-scale computational modeling; neuroscientists will see how their datasets are being used; and policymakers will find a consolidated set of issues that cut across AI governance and brain data regulation.

**PART 1. THE EMERGING LANDSCAPE OF BRAIN FOUNDATION MODELS**

Efforts to understand the human brain have long faced difficulties when it comes to analyzing brain data. The human brain is an extraordinarily complex system; at any moment, there are roughly 86 billion neurons and 100 trillion synapses interacting across different brain regions, varying not only within a person but also across people.[10] Faced with this complexity, most progress in neuroscience and neurotechnology has come from research focusing on a particular brain region, a specific behavior or task, or a single medical concern. While AI has accelerated advancements across all of these, such efforts remain expensive and time consuming, and their

---

analyzing the legal and regulatory gaps for both and contrasting them as unique relative to biometric and biological data. *See* R.I. Field, *The Data We Leave Behind: Limits of Legal Protections for Neurotechnology and Genomic Data*, 15 DREXEL L. REV. 769 (2023).

[10] E.R. Kandel et al., PRINCIPLES OF NEURAL SCIENCE (6th ed. 2021).

results are difficult to scale. It is in this context that brain foundation models are emerging. Rather than being built for a single brain region or a narrowly defined purpose, these models allow for a generalizable and adaptable representation of the brain.

In this section, we provide a descriptive account of these models, including 1) what these models are and which domains and tasks they are likely to be applied to and (2) what training datasets are being used, where they come from, and who creates and uses them.

**1.1. What Are Brain Foundation Models, and How Do They Work?**

Brain foundation models are a departure from the traditional single-purpose AI models trained on brain data over the past decade. These older models have generally been designed for one narrow task and are trained to perform it with a technique called *supervised learning*. We refer to these models throughout the paper as task-specific models. Here, developers start by choosing what they want their model to do, such as detect an epileptic seizure,[11] diagnose schizophrenia,[12] or classify a person's emotional state.[13] Then they train the model with data that have been labeled specifically for that task. If the goal is for the model to detect a seizure, each segment of the training data is labeled "seizure" or "non-seizure." If the goal is to detect schizophrenia, each segment is labeled "schizophrenia" or "non-schizophrenia." In this way, the model learns to identify certain patterns of brain activity with the labels assigned to the data.

Supervised, task-specific models continue to drive major breakthroughs;[14] however, they are costly, time intensive, and difficult to develop. For each new task, developers must find or create domain-specific datasets that contain the phenomena of interest, but these datasets can be inaccessible, especially for certain types of tasks, like diagnosing rare diseases. For example, Creutzfeldt-Jakob Disease is a neurological disease that affects one in one million people a year.[15] It would be extremely difficult to compile a dataset sufficient to build a task-specific model for diagnosis. Labeling requires expert annotation, and neurologists or PhD researchers may spend days reviewing recordings to extract only seconds or minutes of relevant activity,[16] and in fact best practice often requires multiple experts to review the same data before reaching a consensus.[17]

---

[11] M.K. Siddiqui et al., *A Review of Epileptic Seizure Detection Using Machine Learning Classifiers*, 7 BRAIN INFORMATICS 5 (2020).
[12] S. Bagherzadeh et al., *Detection of Schizophrenia Using Hybrid of Deep Learning and Brain Effective Connectivity Image from Electroencephalogram Signal*, 146 COMPUTERS IN BIOLOGY & MEDICINE 105570 (2022).
[13] T.D. Ly & G.H. Ngo, *Consumer-Friendly EEG-Based Emotion Recognition System: A Multi-Scale Convolutional Neural Network Approach*, arXiv (2025), https://arxiv.org/abs/2506.16448.
[14] J. Tang et al., *Semantic Reconstruction of Continuous Language from Non-Invasive Brain Recordings*, 26 NATURE NEUROSCIENCE 858 (2023); O. Sen et al., *A Low-Latency Neural Inference Framework for Real-Time Handwriting Recognition from EEG Signals on an Edge Device*, 15 SCIENTIFIC REPORTS 41040 (2025).

Brain foundation models, in contrast, are trained using *self-supervised* learning, which means that they are exposed to vast datasets of unlabeled brain data from a range of individuals and devices. These models are not trained to map brain activity to labels but to use training data to make simple predictions. Developers mask a short portion of long EEG or fMRI data streams and task the model with predicting the missing section based on what came before and after. By solving these kinds of puzzles tens of thousands of times, the model learns the regularities of neural activity without ever being told "this is a seizure" or "this is schizophrenia." This process of training a model on large unlabeled datasets is called *pretraining* and is a hallmark of the foundation model paradigm.

Through pretraining, the model builds up a generalizable representation of how brain signals typically behave and what the common patterns are across different brain regions,[18] analogous to what LLMs do when they learn grammar and word relationships. These representations are highly adaptable, allowing for rapid calibration and personalization for individuals, as well as transfer across contexts and settings. Furthermore, they are able to be adapted across individuals and contexts with a high degree of data efficiency, such as the ability to perform new tasks with only a few labeled examples, called *few-shot learning*, or in a new context with no additional training, called *zero-shot transfer*. Early results suggest that brain foundation models can accomplish all of this at no cost to accuracy, demonstrating performance on par with and in some cases better than task-specific models.¹⁹ In contrast, traditional supervised machine learning models require large, task-specific labeled datasets, making zero-shot transfer impossible as they cannot generalize beyond what they were explicitly trained on. Few-shot learning is also challenging since neural data varies from person to person, is highly context-sensitive, and tends to be noisy. For these reasons, traditional supervised approaches typically require extensive recalibration or even complete retraining to remain accurate across different users or contexts.

---

[15] L.L. Rong et al., *Five Creutzfeldt-Jakob Disease Cases Within One Year in Spectrum Health Hospitals in Grand Rapids, Michigan*, 100 NEUROLOGY 2527 (2023).
[16] F.A. Nascimento et al., *Competency-Based EEG Education: An Online Routine EEG Examination for Adult and Child Neurology Residents*, 2 NEUROLOGY EDUCATION e200094 (2023).
[17] C. da Silva Lourenço et al., *Efficient Use of Clinical EEG Data for Deep Learning in Epilepsy*, 132 CLINICAL NEUROPHYSIOLOGY 1234 (2021).
[18] J.O. Caro et al., *BrainLM: A Foundation Model for Brain Activity Recordings*, bioRxiv (2023), https://doi.org/10.1101/2023.09.12.557460.

A rapidly growing body of technical work, alongside initial industry projects and pilots, suggests that using the foundation model paradigm on brain data could be adapted into valuable use cases across a range of different domains and applications. In the next section we highlight a few such areas.

## 1.2. What Can Brain Foundation Models Do?: Four Downstream Contexts

In the coming years, what types of medical and nonmedical applications are we likely to see from these models? Much as law firms have fine-tuned LLMs to review and redline contracts,²⁰ tech companies to serve as engineering copilots,²¹ and scientists to assist in drug development,²² brain foundation models are likely to be deployed within a range of downstream contexts and systems.²³ In this section, we present four: personalized medicine and clinical care, neurotechnology systems, basic neuroscience research, and AI training.

---

[19] J. Wu et al., *supra* note 5; Caro et al., *supra* note 18.
[20] *Harvey*, OPENAI (2024), https://openai.com/index/harvey/.
[21] K. Wernet Miller, *Embracing Generative AI for Law*, IRONCLAD (June 11, 2025), https://ironcladapp.com/journal/legal-ai/generative-ai-for-law ; T. Smith, *OpenAI's GPT-5-Codex: A Smarter Approach to Enterprise Development*, DEVOPS.COM (Sept. 17, 2025), https://devops.com/openais-gpt-5-codex-a-smarter-approach-to-enterprise-development/.
[22] X. Liu et al., *Application of Artificial Intelligence Large Language Models in Drug Target Discovery*, 16 FRONTIERS IN PHARMACOLOGY 1597351 (2025).
[23] We highlight these four contexts because they are already represented in research and are attracting early industry attention; however, their real-world effectiveness and reliability remain to be seen. As experience with LLMs suggests, foundation models can be error-prone (including hallucinations), often lack robust mechanistic grounding, and may perform unevenly across tasks and deployment contexts. Brain foundation models are likely to inherit at least some of these issues, though which and to what degree remain unknown. See S. Banerjee et al., *LLMs Will Always Hallucinate, and We Need to Live with This*, in INTELLIGENT SYSTEMS AND APPLICATIONS 624 (Kohei Arai ed., 2025); F. Dell'Acqua et al., *Navigating the Jagged Technological Frontier: Field Experimental Evidence of the Effects of AI on Knowledge Worker Productivity and Quality*, Harv. Bus. Sch. Tech. & Operations Mgmt. Unit Working Paper No. 24-013 (2023).

*1.2.1. Personalized Medicine and Clinical Care*

Around the world, neurological and psychiatric conditions afflict billions of people,[24] with many conditions going undiagnosed due to infrequent checkups and inaccessible diagnostic tools.[25] As populations age, brain health will become a more urgent public health priority,[26] but early research shows that brain foundation models could help. For example, they have demonstrated the ability to diagnose neurological conditions including Alzheimer's disease from brain signals,[27] showing potential for biomarker discovery and future clinical applications.[28] They can also help developing neural digital twins, which are dynamic models of an individual's brain patterns, behaviors, and bodily responses, could allow for individual simulation of a person's disease progression and their treatments before making decisions in the real world.[29]

One of the powerful characteristics of brain foundation models, as described above, is their ability to be fine-tuned for each task with limited data, which is extremely valuable in clinical diagnostics. The model BrainWave, for example, needed to be fine-tuned on only five examples to accurately detect early-stage Alzheimer's, and it needed less than 10 to classify children with Attention-Deficit/Hyperactivity Disorder (ADHD) versus those more typically developing.[30] BrainGFM, trained on MRI data, needed as little as 1% of the full training data, to effectively adapt to new neurological and psychiatric disorders, such as ADHD and Alzheimer's.[31]

BrainWave also demonstrated how zero-shot transfer capabilities of brain foundation models could enable a kind of plug-and-play adoption across clinics and hospitals. When fine-tuned with seizure-detection data from St. Anne's University Hospital Brno in the Czech Republic, it worked with no additional training at the Mayo Clinic, a hospital with different equipment and patient populations.[32]

---

[24] World Health Org., *Over 1 in 3 People Affected by Neurological Conditions, the Leading Cause of Illness and Disability Worldwide* (Mar. 14, 2024), https://www.who.int/news/item/14-03-2024-over-1-in-3-people-affected-by-neurological-conditions--the-leading-cause-of-illness-and-disability-worldwide
[25] A. Bradford et al., *Diagnostic Error in Mental Health: A Review*, 33 BMJ QUALITY & SAFETY 663 (2024).
[26] World Health Org., *Optimizing brain health across the life course: WHO position paper* (2022), https://www.who.int/publications/i/item/9789240054561
[27] Z. Yuan et al., *BrainWave: A Brain Signal Foundation Model for Clinical Applications*, arXiv (2024), https://arxiv.org/abs/2402.10251; Caro et al., *supra* note 18.
[28] Caro et al., *supra* note 18; Zhou et al., *supra* note 4.
[29] L.S. Fekonja et al., *The Digital Twin in Neuroscience: From Theory to Tailored Therapy*, 18 FRONTIERS IN NEUROSCIENCE 1454856 (2024); E.Y. Wang et al., *Foundation Model of Neural Activity Predicts Response to New Stimulus Types*, 640 NATURE 470 (2025).
[30] Yuan et al., *supra* note 27.
[31] X. Wei et al., *A Brain Graph Foundation Model: Pretraining and Prompt-Tuning for Any Atlas and Disorder*, arXiv (2025), https://arxiv.org/abs/2506.02044.

Already, early commercial efforts are underway to integrate models into clinical settings, including an AI start-up that is piloting a system with a major hospital to support neurologists in intensive care units[33] and another using foundation models to forecast medical treatment response.[34]

*1.2.2. Neurotechnology Systems*

A second downstream setting is the integration of brain foundation models into neurotechnologies to enable faster, more accurate, and more intuitive systems. Brain−computer interfaces (BCIs), primarily for people with sensory or motor impairments, have significantly advanced in recent years, now able to decode intended movements like typing[35] or speech[36] at rates and levels of accuracy previously unimaginable. However, these systems require extensive individualized training to tailor the decoding software to each user's neural patterns. For example, a recent system required 16 hours for patient-specific calibration.[37] Brain foundation models could allow faster calibration, possibly within minutes, and more accurate decoding,[38] and they offer an array of features that could enrich users' lives, such as allowing broader forms of creative expression.[39] These models could feasibly contribute to the goal of moving BCI beyond purely therapeutic applications to cognitive enhancement more broadly.[40]

---

[32] Yuan et al., *supra* note 27; Caro et al., *supra* note 18.
[33] Mullin, *supra* note 6
[34] *Foundational Brain Model The Engine Behind Our Research*, BRAINIFY.AI (2025), https://brainify.ai/foundationalmodel.
[35] F.R. Willett et al., *High-Performance Brain-to-Text Communication via Handwriting*, 593 NATURE 249 (2021).
[36] E.M. Kunz et al., *Inner Speech in Motor Cortex and Implications for Speech Neuroprostheses*, 188 CELL 4658 (2025).
[37] Tang et al., *supra* note 14.
[38] D. Liu et al., *MIRepNet: A Pipeline and Foundation Model for EEG-Based Motor Imagery Classification*, arXiv (2025), https://arxiv.org/abs/2507.20254.
[39] P. Zhou et al., *Neural-Driven Image Editing*, arXiv (2025), https://arxiv.org/abs/2507.05397.
[40] E.C. Gordon & A.K. Seth, *Ethical Considerations for the Use of Brain–Computer Interfaces for Cognitive Enhancement*, 22 PLOS BIOLOGY e3002899 (2024).

Beyond clinical grade BCIs, neurotechnology wearables, like EEG headbands or electromyography (EMG) bracelets, also stand to benefit from the integration of brain foundation models. The success of non-neurotech wearables, such as the Oura Ring and Apple Watch, has demonstrated consumer appetite for wearable health tools.[41] The proposed value-add of wearable neurotech is that brain data could enable more accurate analysis for clinical or wellness measures, from detecting stress levels to measuring focus over the workday, and they could even afford novel modes of human−computer interaction, like hands-free typing.[42]

However, the market for these devices has struggled to perform well given how noisy the data are; the sensors sit on the face, inside the ear, or on the scalp and are therefore highly sensitive to jaw or facial movements and hair and scalp contact.[43] Not to mention the high costs associated with building a task-specific model from labeled data. Brain foundation models could hypothetically improve both. In terms of low data quality from hardware, these models are able to separate the noisiness of a user's movement from the actual brain patterns, better than traditional task-specific models.[44] This means they could lead to more accurate and reliable wearable neurotech, a boon for a market that has long struggled with noisy, low quality data and as a result persistent skepticism about their effectiveness.[45]

Second, these models can significantly reduce the cost of developing and releasing new features and applications. Neurotech companies that originally released a headband for sleep could prototype, test, and deploy features for mood monitoring, meditation, or focus management for work, all for a fraction of previous costs. Also, in the same way that numerous start-ups or established companies have been able to use GPT models to build apps or new features, these models can lower the barrier to entry for small organizations without large in-house data or machine-learning resources.

---

[41] L. Piwek et al., *The Rise of Consumer Health Wearables: Promises and Barriers*, 13 PLOS MEDICINE e1001953 (2016).
[42] P. Kaifosh & T.R. Reardon, *A Generic Non-Invasive Neuromotor Interface for Human-Computer Interaction*, 645 NATURE 702 (2025).
[43] M.K. Islam et al., *Methods for Artifact Detection and Removal from Scalp EEG: A Review*, 46 NEUROPHYSIOLOGIE CLINIQUE 287 (2016).
[44] N.M. Foumani et al., *EEG-X: Device-Agnostic and Noise-Robust Foundation Model for EEG*, arXiv (2025), https://arxiv.org/abs/2511.08861.
[45] A. Wexler, *Separating Neuroethics from Neurohype*, 37 NATURE BIOTECHNOLOGY 988 (2019).

Finally, they could allow for the introduction of clinical tools where they have been prohibitively expensive or simply unavailable. Returning to the example of Creutzfeldt-Jakob Disease, building a task-specific algorithm is almost impossible for a single institution. Practitioners might have to spend 10 years collecting around 100 cases and undertake expensive reporting for those patients, including chronic, long-term 24-hour recording, just to train an acceptable algorithm. A brain foundation model deployed on a consumer-grade neurotech device could make the diagnosis of this and other rare diseases globally available in communities with little medical access.

We are seeing the initial wave of neurotech companies developing and integrating models with their hardware. At least one invasive BCI company has announced an in-house brain foundation model intended for deployment alongside its implant system,[46] and several direct-to-consumer wearable neurotechnology companies have publicly mentioned developing proprietary and open-source brain foundation models,[47] with one integrating the model into its proprietary headbands for clinical-grade sleep monitoring.[48]

*1.2.3. Basic Neuroscience Research*

The third downstream setting is basic neuroscience research, a field long constrained by the slow pace and high costs of conducting studies. On top of costly and labor-intensive components of supervised learning (detailed above), the availability of data in neuroscience has been a bottleneck to conducting research,[49] and experiments take years to run, analyze, and validate.[50]

Neuroscientists point to brain foundation models as a means of reducing the cost and time required to generate new hypotheses and run experiments.[51] Instead of each individual lab training its own model for a single experiment, researchers could plug their own data into a model many studies have shared. This shared infrastructure of datasets, tools, and models could allow scientists to test whether patterns first observed in a tightly controlled task also appear in other datasets, or it could help them predict the outcomes of proposed experiments and prioritize the most informative ones before committing to expensive, multiyear studies.[52]

---

[46] Conor Hale, *Synchron and Nvidia Set Sights on AI Model Trained by Direct Brain Activity*, FIERCE BIOTECH (Mar. 19, 2025), https://www.fiercebiotech.com/medtech/synchron-and-nvidia-set-sights-ai-model-trained-direct-brain-activity.
[47] J. Nixon & A.J. Keller, Neurosity EEG Dataset, GITHUB (2024), https://github.com/JeremyNixon/neurosity.
[48] *Muse's AI Platform*, *supra* note 7.
[49] S. Yang et al., *Foundation and Large-Scale AI Models in Neuroscience: A Comprehensive Review*, arXiv (2025), https://arxiv.org/abs/2510.16658.
[50] E. Dyer & B. Richards, *Accepting "the Bitter Lesson" and Embracing the Brain's Complexity*, THE TRANSMITTER (Mar. 26, 2025), https://www.thetransmitter.org/neuroai/accepting-the-bitter-lesson-and-embracing-the-brains-complexity/.
[51] Caro et al., *supra* note 18.

Beyond this, models could become a backbone for multiscale efforts to model the brain and understand the electrical and biological principles that underlie its activity.[53] More and more projects aim to collect, perturb, and map neural activity across different modalities (e.g., from macroscopic imaging data like MRI and positron emission tomography to microscopic cellular modeling using electron microscopy reconstruction) and generate digital simulations of brain activity.[54] In this type of multidisciplinary project, brain foundation models could train not only on different types of brain data, like EEG or fMRI, but also different modalities, allowing researchers to explore the brain's functioning across domains.[55]

*1.2.4. Training and Improving AI Systems*

A growing body of research asks whether the brain or the use of brain data can correct limitations of current AI systems. Today's LLMs often exhibit a "jagged frontier" of capability, where impressive reasoning strengths coexist with surprising brittleness.[56] Recent evaluations using the ConceptARC benchmark make this unevenness concrete: several models appeared to match human accuracy on text-based logic tasks, but their internal "explanations" often revealed a reliance on superficial statistical shortcuts rather than true abstraction. Furthermore,

---

[52] Dyer & Richards, *supra* note 50.
[53] Dyer & Richards, *supra* note 50.
[54] MICrONS Consortium, *Functional Connectomics Spanning Multiple Areas of Mouse Visual Cortex*, 640 NATURE 435 (2025); H. Xiong et al., *Digital Twin Brain: A Simulation and Assimilation Platform for Whole Human Brain*, arXiv (2023), https://arxiv.org/abs/2308.01941.
[55] G. Haspel et al., *The Time Is Ripe to Reverse Engineer an Entire Nervous System: Simulating Behavior from Neural Interactions*, arXiv (2023), https://arxiv.org/abs/2308.06578.
[56] F. Dell'Acqua et al., Navigating the Jagged Technological Frontier: Field Experimental Evidence of the Effects of AI on Knowledge Worker Productivity and Quality, Harv. Bus. Sch. Tech. & Operations Mgmt. Unit Working Paper No. 24-013 (2023)

performance dropped sharply when the same tasks were presented visually.[57] These results suggest that current systems do not yet reliably form or apply abstractions the way humans do.

Researchers have recently turned to neural data as a form of process supervision to guide how a model computes, rather than just what it predicts. By training a model to align its latent representations with human neural activity during perception or reasoning—coined in recent work as Brain-Informed Process Supervision (BIPS)—researchers can "inject" biological robustness into artificial systems.[58] A study showed that by "brain-tuning" speech model to match fMRI response, the data efficiency and generalizability of the model were significantly increased.[59] This work overlaps with a broader push by many technologists for the AI field to go beyond "bigger models and next-token prediction" towards architectures that borrow organizing principles from neural circuits often termed "neuromorphic" or brain-inspired design. This technique has been shown to significantly improve image recognition stability[60] and increase computational efficiency.[61]

While these efforts are not always "brain foundation model training" in the strict sense, they illustrate a cohesive NeuroAI program: brain data serves as a form of supervision to guide learning, while brain-inspired designs shape the architecture to favor robust generalization. This community also increasingly views these techniques as vital for evaluating AI safety, as aligning a model's internal processes with human-like reasoning may reduce the risk of "shortcut learning" in critical systems.[62]

---

[57] C. Beger et al., *Do AI Models Perform Human-like Abstract Reasoning Across Modalities?,* arXiv (2025), https://arxiv.org/abs/2510.02125.
[58] P. Mineault et al., *NeuroAI for AI Safety*, arXiv (2024), https://arxiv.org/abs/2411.18526
[59] O. Moussa et al., *Improving Semantic Understanding in Speech Language Models via Brain-Tuning*, arXiv (2024), https://arxiv.org/abs/2410.09230.
[60] J. Dapello et al., *Simulating a Primary Visual Cortex at the Front of CNNs Improves Robustness to Image Perturbations*, in ADVANCES IN NEURAL INFORMATION PROCESSING SYSTEMS 33 (2020).
[61] R. Moyal et al., *Heterogeneous Quantization Regularizes Spiking Neural Network Activity*, 15 SCIENTIFIC REPORTS 14045 (2025).
[62] Y. LeCun, (2022)., *A Path Towards Autonomous Machine Intelligence*. OPENREVIEW *(2022)*, https://openreview.net/pdf?id=BZ5a1r-kVsf; D. Hassabis et al., *Neuroscience-Inspired Artificial Intelligence*, 95 NEURON 245 (2017); Mineault et al., *supra* note 58.

Importantly, this downstream context reframes brain data as a general AI training resource, likely intensifying incentives to scale collection beyond traditional clinical and research boundaries—and raising the stakes of upstream training-data governance.

In highlighting these four settings, we have tried to concretely show how brain foundation models might be integrated into different domains and systems as well as the kinds of benefits they could offer individuals and society at large. All of these developments, however, rest on data; a single brain foundation model requires training on tens of thousands of hours of unlabeled brain data captured on multiple devices from diverse hardware and participant groups.[63] Next we turn to the training data ecosystem underpinning the development of these models.

### 1.3. Training Data Ecosystem

Unlike the text and images scraped from the public internet to fuel LLMs and other generative AI systems, there is no equivalent open web of brain data that researchers can use to train models. Even so, current state-of-the-art brain foundation models are trained on datasets that are orders of magnitude larger than those used for most task-specific models in neuroscience. For example, a traditional epilepsy-detection model might be trained on roughly 80 hours of EEG recordings,[64] whereas BrainWave was trained on over 40,000 hours of EEG and invasive EEG recordings,[65] and BrainLM was trained on 6,700 hours of fMRI scans.[66] Where, therefore, are developers getting the requisite volumes of brain data?

Today, brain foundation model developers draw primarily from two broad sources. The first consists of publicly available clinical and research datasets originating in clinical care and academic studies. The second consists of proprietary datasets generated by private firms, including consumer neurotechnology companies, invasive brain computer interface companies, and large technology firms. In this section, we will describe each of these sources, and then consider dynamics introduced by these data sources as training data.

---

[63] Yuan et al., *supra* note 27.
[64] Y. Berrich & Z.E. Guennoun, *EEG-Based Epilepsy Detection Using CNN-SVM and DNN-SVM with Feature Dimensionality Reduction by PCA*, 15 SCIENTIFIC REPORTS 14313 (2025).
[65] Yuan et al., *supra* note 27.
[66] Caro et al., *supra* note 18.

*1.3.1. Public Research/Clinical Datasets*

Publicly available datasets form the training data bedrock for brain foundation models. Historically, collecting brain data has required specialized sensors and hardware and has taken place primarily in academic studies, clinical trials, or routine clinical care. Many of the resulting datasets were funded by government agencies, such as the U.S. National Institutes of Health. Under an open-science model and ethos,[67] these datasets are anonymized and made available to the broader research community, reflecting a long-standing commitment in neuroscience to data sharing. Archives such as OpenNeuro, the Distributed Archives for Neurophysiology Data Integration (DANDI), the Temple University Hospital EEG Data Corpus, and the Data Archive for the BRAIN Initiative (DABI) house hundreds of openly available datasets.[68]

Long before brain foundation models, public datasets were already central to neuroscience research. Researchers used them to develop and benchmark task-specific AI models,[69] students drew on them for coursework and competitions,[70] and amateur neurotechnologists used them, sometimes via community EEG platforms, to build small applications.[71]

Now they are being repurposed to develop brain foundation models.[72] This is a novel development, and it is due to the fact that brain foundation models are able to make use of these archives of large, heterogeneous datasets. Unlike task-specific systems, they can absorb data without requiring that every recording be accompanied by task-specific labels or even collected under a single protocol. Further, because brain foundation models are able to train on multiple kinds of brain data in a single system (unlike task-specific models), one model can combine datasets with different kinds of data, like MRI and fMRI scans, which would previously have been extremely difficult if not impossible.[73] Experimental systems go further by pairing brain data with text. In one model, brain data is trained with electronic health records, so the model also learns how patterns of neural activity relate to the behavioral and clinical concepts used to describe them.[74]

---

[67] I. Obeid & J. Picone, *The Temple University Hospital EEG Data Corpus*, 10 FRONTIERS IN NEUROSCIENCE 196 (2016); K.J. Gorgolewski et al., *OpenNeuro: A Free and Open Platform for Sharing and Analyzing Neuroimaging Data*, 6 ELIFE e27712 (2017); Y.O. Halchenko et al., *DANDI: A Platform for Publishing and Sharing Neurophysiology Data*, 18 NATURE METHODS 1138 (2021).
[68] D. Duncan et al., *Data Archive for the BRAIN Initiative (DABI)*, 10 SCIENTIFIC DATA 83 (2023).
[69] C.J. Markiewicz et al., *The OpenNeuro Resource for Sharing of Neuroscience Data*, 10 ELIFE e71774 (2021); DANDI Archive (RRID SCR_017571), https://dandiarchive.org; Neural Engineering Data Consortium, *The Neural Engineering Data Consortium v5.0* (May 2, 2024), https://www.nedcdata.org/.
[70] J.A. Segawa, *Hands-On Undergraduate Experiences Using Low-Cost Electroencephalography (EEG) Devices*, 30 J. UNDERGRADUATE NEUROSCIENCE EDUCATION A11 (2019).
[71] OpenBCI, https://openbci.com/.
[72] Zhou et al., *supra* note 4.

The Temple University Hospital EEG corpus illustrates this repurposing. It includes seizure recordings from over 14,000 patients collected in the early 2000s,[75] and these were used to train several of the frontier brain foundation models.[76] Similarly, the Alzheimer's Disease Neuroimaging Initiative dataset, constituting longitudinal MRI scans from over 1,700 participants with various degrees of cognitive impairment,[77] was used to train foundation models, such as the BrainIAC.[78]

*1.3.2. Commercial/Proprietary Datasets*

Alongside this public infrastructure, there is a rapidly growing base of commercial actors generating brain data and creating brain datasets. Consumer neurotechnology devices alone collect neural recordings from millions of users daily.[79] Big tech is entering the space: Meta's EMG wristband collects brain data,[80] and while Apple has not publicly released plans, it patented EEG-sensing AirPods in 2023.[81] At the same time, invasive BCI companies, such as Neuralink, Paradromics, and Synchron, are streaming continuous data from patients with implanted devices.

---

[73] Z. Dong et al., *Brain Harmony: A Multimodal Foundation Model Unifying Morphology and Function into 1D Tokens*, arXiv (2025), https://arxiv.org/abs/2509.24693.
[74] S. Gijsen & K. Ritter, *EEG-Language Modeling for Pathology Detection*, arXiv (2024), https://arxiv.org/abs/2409.07480.
[75] Obeid & Picone, *supra* note 67.
[76] Yuan et al., *supra* note 27.
[77] R.C. Petersen et al., *Alzheimer's Disease Neuroimaging Initiative (ADNI) Clinical Characterization*, 74 NEUROLOGY 201 (2010).
[78] D. Tak et al., *A Foundation Model for Generalized Brain MRI Analysis*, medRxiv (2024), https://www.medrxiv.org/content/10.1101/2024.12.02.24317992v1.
[79] N. A. Farahany, THE BATTLE FOR YOUR BRAIN: DEFENDING THE RIGHT TO THINK FREELY IN THE AGE OF NEUROTECHNOLOGY (2023); J. Sabio et al., *A Scoping Review on the Use of Consumer-Grade EEG Devices for Research*, 19 PLOS ONE e0291186 (2024).
[80] D. Sussillo et al., *A Generic Noninvasive Neuromotor Interface for Human-Computer Interaction*, bioRxiv (2024), https://www.biorxiv.org/content/10.1101/2024.02.23.581779v1.
[81] Erdrin Azemi et al., *Biosignal Sensing Device Using Dynamic Selection of Electrodes*, U.S. Patent Application Publication No. 2023/0225659 A1 (filed Jan. 9, 2023).

While these companies currently market primarily to people with sensory and motor disorders, some openly describe their long-term ambitions to introduce BCIs to consumer markets.[82]

As these firms usher in a major influx of proprietary neural data, the sector is likely to see the emergence of a data market. In this model, the value proposition shifts: a company may collect neural data not to improve its own hardware or internal models, but specifically to sell or license that data as a commodity to third-party developers. We are already seeing the first iterations of this in "Brain-Data-as-a-Service" (BDaaS) models:[83] providers that standardize and license neural data for others to build "Large Mental Models."

This trajectory, where data moves from the device manufacturer to a secondary buyer, is particularly likely in the early days of the sector as firms search for durable revenue streams. Absent explicit constraints, both consumer and invasive neurotech companies may be incentivized to monetize their datasets via licensing or sale to interested companies. This would mirror the mature "information reseller" ecosystem in other consumer-tech domains, where data brokers aggregate and trade personal data as a primary business model.[84]

*1.3.3. Training Data Ecosystem: Stitching and Scaling Dynamics*

Taken together, these public and proprietary datasets now supply training data for brain foundation models developed across universities and commercial actors alike, including invasive BCI companies,[85] big tech,[86] neurotechnology start-ups, and software-only model developers.[87] Further, these developers do not train on the various sources in isolation. In practice, they stitch together datasets from across contexts, combining clinical recordings, consumer-device EEGs, and decades-old seizure records from an academic research study.

---

[82] L. Leffer, Neuralink's First User Describes Life with Elon Musk's Brain Chip, SCIENTIFIC AMERICAN (June 7, 2024), https://www.scientificamerican.com/article/neuralinks-first-user-describes-life-with-elon-musks-brain-chip/ ; K. Choudhury, *Neuralink Targets $1 Billion Revenue by 2031, Bloomberg News Reports*, REUTERS (July 3, 2025), https://www.reuters.com/technology/neuralink-targets-1-billion-revenue-by-2031-bloomberg-news-reports-2025-07-23/.
[83] A. Vecchione, *Brain.space Secures $11M to Grow Its Mental Model Platform*, MOBIHEALTHNEWS (March 18, 2025), https://www.mobihealthnews.com/news/brainspace-secures-11m-grow-its-mental-model-platform.
[84] Federal Trade Commission, *Data Brokers: A Call for Transparency and Accountability* (May 2014), https://www.ftc.gov/reports/data-brokers-call-transparency-accountability-report-federal-trade-commission-may-2014.
[85] Hale, *supra* note 46.
[86] Meta, *Human-Computer Input via a Wrist-Based sEMG Wearable*, META BLOG (January 9, 2025), https://www.meta.com/blog/surface-emg-wrist-white-paper-reality-labs/.
[87] Mullin, *supra* note 6.

This means that private, for-profit companies can pretrain proprietary brain foundation models on a mixture of publicly funded legacy datasets and data from their own users; for example, Emotiv, a neurotechnology company specializing in neuromarketing and consumer wellness, has trained a foundation model on the Temple University Hospital EEG Data Corpus as well as its own proprietary data.[88] Neurosity, another neurotechnology start-up, has open-sourced a brain foundation model it built using Temple data likely supplemented with data from its own device.[89] These cases are early examples of an emerging pattern in which brain foundation models collapse data from distinct contexts into a shared substrate.

These dynamics will only accelerate. Like other data-hungry areas of AI, these models exhibit *scaling law* behavior, in which model performance improves as models are trained on larger datasets, and current systems show no signs of having exhausted these gains.[90] Beyond more data, the field is also motivated to seek out more diverse data, as model performance improves with training on data collected from diverse contexts.[91] For example, consumer wearable data is valuable because it captures continuous, ambient recordings of brain activity during everyday activities, like typing, viewing content, and interacting socially, which are noisy but real-world inputs that clinical settings cannot replicate.[92] In short, demand for brain data of all different types will not plateau.[93]

---

[88] J. Hong et al., *EEGM2: An Efficient Mamba-2-Based Self-Supervised Framework for Long-Sequence EEG Modeling*, arXiv (2025), https://arxiv.org/abs/2502.17873.
[89] Nixon & Keller, *supra* note 47.
[90] Wang et al. and Jayalath et al. each reported monotonic accuracy gains as pretraining hours grow. Wang et al., *supra* note 29; D. Jayalath et al., *The Brain's Bitter Lesson: Scaling Speech Decoding With Self-Supervised Learning*, arXiv (2025), https://arxiv.org/abs/2406.04328. Jiang et al. showed LaBraM still improving beyond 10,000 hours with no sign of saturation. W. Jiang et al., *Large Brain Model for Learning Generic Representations with Tremendous EEG Data in BCI*, in INT'L CONF. ON LEARNING REPRESENTATIONS (ICLR) (2024). This mirrors the scaling laws first charted for language models. J. Kaplan et al., Scaling Laws for Neural Language Models, arXiv (2020), https://arxiv.org/abs/2001.08361. In a neuro-to-LLM comparison, the datasets currently used for pretraining brain foundation models were considerably smaller than those used for state-of-the-art LLMs, which can involve tens of trillions of tokens. See D. Kostas et al., *BENDR: Using Transformers and a Contrastive Self-Supervised Learning Task to Learn from Massive Amounts of EEG Data*, 15 FRONTIERS IN HUMAN NEUROSCIENCE 653659 (2021).; Bommasani et al., *supra* note 2.
[91] Yuan et al., *supra* note 27.

Furthermore, the vision described in section 1.2.4 of scaled neural data to train more general-purpose AI, may further accelerate collection, as it recasts it as a general AI training resource. Even if the ultimate ambition is debated, it is already influencing research and funding priorities, and it may intensify pressures to scale data through commercial devices and data markets.

Taken together, these trends all point towards a complex mixture and rapidly expanding ecosystem of brain data from more sources, stitched into large training datasets and reused across an expanding set of applications. In Part 2, we turn from a descriptive account to the normative questions this ecosystem raises.

## PART 2: ETHICAL, SOCIAL, AND LEGAL QUESTIONS FOR BRAIN FOUNDATION MODEL TRAINING DATA

Brain foundation models convert brain data—historically governed through clinical care and research ethics—into a reusable training substrate for multi-purpose AI systems. That shift creates a distinctive upstream training environment: datasets are repurposed, stitched across contexts, and reused across downstream domains in ways that existing governance regimes were not built to track.

Part 2 maps the ethical, social, and legal questions that follow from this intersection. Drawing on emerging technical practice, the foundation-model governance literature, and extensive scholarship in neuroethics and bioethics, we organize concerns across five familiar domains: privacy, consent, bias, benefit sharing, and governance. Some issues in these domains are recognizable—variants of problems already documented in AI systems and in biomedical and neuroscience data practice—though they may take on new force when the underlying data are neural. Others arise more specifically from training general-purpose models on brain data and may demand new standards, new techniques, and new institutional arrangements.

---

[92] A. Biondi et al., *Noninvasive Mobile EEG as a Tool for Seizure Monitoring and Management: A Systematic Review*, 63 EPILEPSIA 1041 (2022).
[93] M. Sato et al., *Scaling Law in Brain Data: Non-Invasive Speech Decoding with 175 Hours of EEG Data*, arXiv (2024), https://arxiv.org/abs/2407.07595; H. Banville et al., *Scaling Laws for Decoding Images from Brain Activity*, arXiv (2025), https://arxiv.org/abs/2501.15322.

We do not attempt to resolve the hardest questions here. Instead, for each domain we surface agenda-setting questions the field will need to confront as the technology matures. At the same time, the field does not need to wait for full resolution before acting. For each domain, we also propose baseline safeguards, practical defaults that can be implemented now to raise the floor for training-data governance while the more difficult questions remain open.

**2.1. Privacy**

In the United States and internationally, privacy concerns are prominent in the neuroethics literature and emerging neurotechnology policy landscape.[94] To date, however, most discussions have focused on the collection, storage, and use of brain data from nonmedical consumer neurotechnology and less on using these data for training foundation AI models or other computational modeling.[95] Work identifying risks in foundation models, in contrast, have highlighted the privacy concerns that arise as a function of this new type of model, and in particular from its training data methods.

*2.1.1. Reidentifying Individuals in Sensitive, Public Brain Datasets*

Making clinical and research datasets public is a cornerstone of open science and is widely seen as essential for advancing neuroscience and medicine.[96] At the same time, this practice creates a well-known privacy tension between maximizing the accessibility and scientific utility of datasets and protecting the privacy of the people whose data they contain.[97] Even when datasets

---

[94] Magee et al., *Beyond Neural Data: Cognitive Biometrics and Mental Privacy*, 112 NEURON (2024).
[95] J. Spivack & C. Victory, *The Neural Data Goldilocks Problem: Defining Neural Data in U.S. State Privacy Laws*, FUTURE OF PRIVACY FORUM (Aug. 12, 2025), https://fpf.org/blog/the-neural-data-goldilocks-problem-defining-neural-data-in-u-s-state-privacy-laws/; United Nations Educational, Scientific and Cultural Organization, *Draft Recommendation on the Ethics of Neurotechnology*, SHS/BIO/REC-NEURO/2025 (2025), https://www.unesco.org/en/ethics-neurotech/recommendation.
[96] R.A. Poldrack & K.J. Gorgolewski, *Making Big Data Open: Data Sharing in Neuroimaging*, 17 NATURE NEUROSCIENCE 1510 (2014); R.A. Poldrack et al., *Scanning the Horizon: Towards Transparent and Reproducible Neuroimaging Research*, 18 NATURE REVIEWS NEUROSCIENCE 115 (2017); T.E. Nichols et al., *Best Practices in Data Analysis and Sharing in Neuroimaging Using MRI*, 20 NATURE NEUROSCIENCE 299 (2017); M.D. Wilkinson et al., *The FAIR Guiding Principles for Scientific Data Management and Stewardship*, 3 SCIENTIFIC DATA 160018 (2016).
[97] P. Ohm, *Broken Promises of Privacy: Responding to the Surprising Failure of Anonymization*, 57 UCLA L. REV. 1701 (2010); D.O. Eke et al., *Pseudonymisation of Neuroimages and Data Protection: Increasing Access to Data*

are released after de-identification or anonymization under frameworks such as the Health Insurance Portability and Accountability Act's (HIPAA) Safe Harbor,[98] research demonstrates that these data can be linked to individuals, especially when those datasets are combined with other sources.[99] When the data encode uniquely identifying information, like biometric data including brain data, DNA, and genomic data, it can act like a fingerprint.[100]

Combining datasets for foundation model training, alongside the proliferation of consumer neurotech, may allow bad actors to identify individuals as members of supposedly de-identified datasets, a risk referred to in technical literature as a *membership inference attack*[101]. In principle, a neurotechnology company could pretrain a brain foundation model on publicly available anonymized clinical epilepsy archives alongside their own proprietary user data from their consumer neurotech headband. If one of their users contributed to these clinical or research datasets, could the company infer that they had participated in an epilepsy study or Alzheimer's clinical trial?

This has not been empirically demonstrated in brain foundation models, but has been observed in neuroimaging models and EEG classification algorithms used in BCI systems, as well as in deep

---

*While Retaining Scientific Utility*, 1 NEUROIMAGE: REPORTS 100053 (2021); T. White et al., *Data Sharing and Privacy Issues in Neuroimaging Research: Opportunities, Obstacles, Challenges, and Monsters Under the Bed*, 43 HUMAN BRAIN MAPPING 278 (2022).
[98] 45 C.F.R. § 164.514(b)(2).
[99] G. Loukides et al., *The Disclosure of Diagnosis Codes Can Breach Research Participants' Privacy*, 17 J. AM. MED. INFORMATICS ASS'N 322 (2010); L. Rocher et al., *Estimating the Success of Re-Identifications in Incomplete Datasets Using Generative Models*, 10 NATURE COMMUNICATIONS 3069 (2019); M.A. Rothstein, *Is Deidentification Sufficient to Protect Health Privacy in Research?*, 10 AM. J. BIOETHICS 3 (2010); B.M. Knoppers et al., *Questioning the Limits of Genomic Privacy*, 91 AM. J. HUM. GENETICS 577 (2012); Eke et al., *supra* note 97; A. Narayanan & V. Shmatikov, *Robust De-Anonymization of Large Sparse Datasets*, *in* 2008 IEEE SYMPOSIUM ON SECURITY AND PRIVACY 111 (2008).
[100] D. Jeswani et al., *A Feasibility Study on Using EEG for Biometric Trait Authentication System*, 9 CURRENT DIRECTIONS IN BIOMEDICAL ENGINEERING (2023); A.B. Tatar, *Biometric Identification System Using EEG Signals*, 35 NEURAL COMPUTING & APPLICATIONS 1009 (2023); D. La Rocca et al., *Human Brain Distinctiveness Based on EEG Spectral Coherence Connectivity*, 61 IEEE TRANSACTIONS ON BIOMEDICAL ENGINEERING 2406 (2014); E.S. Finn et al., *Functional Connectome Fingerprinting: Identifying Individuals Using Patterns of Brain Connectivity*, 18 NATURE NEUROSCIENCE 1664 (2015); B. da Silva Castanheira et al., *Brief Segments of Neurophysiological Activity Enable Individual Differentiation*, 12 NATURE COMMUNICATIONS 5713 (2021); Presidential Commission for the Study of Bioethical Issues, *Privacy and Progress in Whole Genome Sequencing* (2012).
[101] R. Shokri et al., *Membership Inference Attacks Against Machine Learning Models*, *in* 2017 IEEE SYMPOSIUM ON SECURITY AND PRIVACY 3 (2017).

learning AI models.[102] Such inferences have proven to translate into discriminatory and harmful decisions in employment, insurance, or credit settings, and could do so here as well.[103]

Finally, this risk may be amplified by current commercial data practices. If consumer neurotechnology companies sell their data, for example to data brokers or third parties, membership inference attacks could be amplified at scale. Currently, many consumer neurotech companies reserve broad rights to share aggregate de-identified brain data with third parties, and some are expressly able to retain data after users delete their accounts.[104]

*2.1.2. Leaking Sensitive Training Data*

Another potential concern relates to whether brain foundation models can leak training data. A growing body of work on risks in foundation models related to training data extraction and memorization shows that models can sometimes reproduce fragments of their training data verbatim when probed in the right way.[105] For example, biomedical researchers have demonstrated the ability to successfully extract medical data used to train or fine-tune LLMs, turning memorization into a channel for leaking protected health information.[106]

Is this possible with brain foundation models? It seems reasonable to assume that models will certainly memorize, even if it is on the basis of brain data. However, whether they do and how a user would go about eliciting the information from a system are both unknowns. While we might be able to imagine how someone could elicit data about an individual from a chatbot with text, it is less obvious how someone might attempt to elicit data about an individual from a brain foundation model; would it require inputting distinctive fragments of an individual's EEG trace to "autocomplete" if a matching brain data prefix is supplied? As models increasingly pretrain on both brain data and text, users may be able to query them in natural language. This raises the possibility that effective prompting could extract information about individuals that was in the pretraining datasets, including rare medical conditions or treatment history.

---

[102] U. Gupta et al., *Membership Inference Attacks on Deep Regression Models for Neuroimaging*, in MEDICAL IMAGING WITH DEEP LEARNING 228 (2021); V. Cobilean et al., *Investigating Membership Inference Attacks Against CNN Models for BCI Systems*, IEEE J. BIOMEDICAL HEALTH INFORMATICS (2025); M. Wu et al., *Evaluation of Inference Attack Models for Deep Learning on Medical Data*, arXiv (2020), https://arxiv.org/abs/2011.00177; M. Duan et al., *Do Membership Inference Attacks Work on Large Language Models?*, arXiv (2024), https://arxiv.org/abs/2402.07841.
[103] Mobley v. Workday, Inc., No. 3:23-cv-00770 (N.D. Cal. filed Feb. 20, 2024); EEOC v. iTutorGroup, Inc., No. 1:22-cv-02565 (E.D.N.Y. settled Aug. 2023); M.M. Mello & S. Rose, *Denial—Artificial Intelligence Tools and Health Insurance Coverage Decisions*, 5 JAMA HEALTH FORUM e240622 (2024); D. K. Citron & F. Pasquale, *The Scored Society: Due Process for Automated Predictions*, 89 WASH. L. REV 1 (2014).
[104] Magee et al., *supra* note 94.
[105] A. Xiong et al., *The Landscape of Memorization in LLMs: Mechanisms, Measurement, and Mitigation*, arXiv (2025), https://arxiv.org/abs/2507.05578.
[106] S. Tonekaboni et al., *An Investigation of Memorization Risk in Healthcare Foundation Models*, arXiv (2025), https://arxiv.org/abs/2510.12950; M. Kim et al., *Fine-Tuning LLMs with Medical Data: Can Safety Be Ensured?*, 2 NEJM AI AIcs2400390 (2025).

*2.1.3. Powerful, Cross-Domain, and Cross-Modality Inferences*

A third category of privacy concerns comes from the unprecedented inferences made possible by brain foundation models. Traditional supervised models require bespoke paired datasets for each relationship under study. Therefore, in order to link for example EEG to emotional states requires a labelled dataset containing both, linking fMRI to behaviour demands another. Brain foundation models address this bottleneck due to their capability to be pretrained on all kinds of data and learn relationships between all of them by mapping disparate data types into a shared representational space.

The novel risk of cross-modality inferences is exemplified by the development of "first-person" foundation models. These models combine neural signals with synchronized streams of high-resolution behavioral and physiological data, such as head-mounted camera footage and galvanic skin response.[107] By ingesting these data, the model is able to learn relationships across different elements of a person's experiences; for example between an individual's internal mental processes and their external sensory environment. This could feasibly allow a model to reconstruct subjective experiences, such as what a person is seeing, feeling, or intending, without the need for extensive and individualized calibration.

The normative concern is twofold. First, these models can draw powerful inferences unthinkable or undetectable by humans. Living under such conditions means accepting completely non-intuitive and unthinkable, while statistically valid, inferences as part of everyday life, casting a disorienting, menacing shadow over even the most mundane behavior.[108] The chilling effect reaches beyond behavior into thought itself. The freedom to think and feel without external judgment depends on some confidence that our internal states remain internal. That confidence disappears when we cannot know which of them are legible to external systems.[109]

---

[107] D. Gamez et al., *A New Type of Foundation Model Based on Recordings of People's Emotions and Physiology*, arXiv (2024), https://arxiv.org/abs/2408.00030; D. Barcari et al., *Recording First-Person Experiences to Build a New Type of Foundation Model*, arXiv (2024), https://arxiv.org/abs/2408.02680.
[108] S. Barocas, *Panic Inducing: Data Mining, Fairness, and Privacy* (Ph.D. dissertation, N.Y. Univ. 2014).

*2.1.4 Agenda-Setting Questions*

Brain foundation models raise privacy questions not only about individual data protection, but about the appropriateness of new information flows across contexts (research, clinical care, consumer devices, and third parties). These flows underpin open science and medical innovation; if trust erodes, it threatens participation and tolerance for data sharing.[110] Against that backdrop, we pose the following critical questions:

1. How can we ensure that people are not re-identified, or linked to sensitive, medical information, through anonymized public data sets? What technical and governance measures need to be adopted or developed?
2. What evaluation methods should developers use to detect and prevent training-data memorization or leakage? Especially as brain data are paired with text or other human-readable modalities?
3. As novel cross-domain inferences become possible from multimodal training, how do we measure how accurate they are and characterize the emergent risks they pose? Should, and if so what, restrictions should be in place regarding what types of data can be combined by what actors?
4. Do the privacy risks specific to brain-model training data justify shifting the default from open release (e.g., open weights) to more controlled release modes (e.g., API access or sandboxed evaluation)?

*2.1.5 Baseline Safeguards*

The harder questions surrounding technical and contextual privacy risks will require time and empirical research to address. But certain baseline safeguards for mitigating privacy risks are available now. Model developers should evaluate datasets for their potential to enable membership inference, memorization, and training-data leakage, a practice that is technically feasible and increasingly standard in the AI field.[111] Where brain foundation models are trained on multimodal data, dataset documentation should specify not only what data is used but the cross-domain inferences the model could enable. We discuss tools for robust dataset provenance documentation, a prerequisite for these evaluations, in the bias section. In the event that model capabilities are uncertain, or when the risks are poorly characterized, developers should default to controlled access (e.g., API, sandbox, or trusted research environment) as opposed to open release, even if the restricted access limits socially beneficial uses.[112] The hope is that, over time, uncertainties can be resolved and tradeoffs minimized, if not reconciled.

---

[109] Farahany, *supra* note 79.
[110] H. Nissenbaum, PRIVACY IN CONTEXT: TECHNOLOGY, POLICY, AND THE INTEGRITY OF SOCIAL LIFE (2010); Baines et al., *Patient and Public Willingness to Share Personal Health Data for Third-Party or Secondary Uses: Systematic Review*, 26 J. MEDICAL INTERNET RESEARCH e50421 (2024).

## 2.2. Consent

In traditional academic research and clinical care, data subjects consent at the point of collection.[113] Ideally, participants are informed about what their data can be used for, how long they can be kept, whether and how they can be shared, and with whom and on what terms. When datasets are subsequently made public or shared to repositories, they are typically accompanied by data-use agreements (DUAs) that set conditions for downstream users. In theory, that process should ensure that consent is carried through to model development and use. In practice, however, consent language and DUAs vary tremendously, from reuse terms that impose restrictions, such as prohibitions against malicious use or re-identification,[114] to very broad, open-ended terms, such as "agree for future research on health and disease," sometimes even ranging to commercial use.[115]

---

[111] Nat'l Inst. of Standards & Tech., Artificial Intelligence Risk Management Framework (AI RMF 1.0), NIST AI 100-1 (January 2023), https://nvlpubs.nist.gov/nistpubs/ai/nist.ai.100-1.pdf; Nat'l Inst. of Standards & Tech., Artificial Intelligence Risk Management Framework: Generative Artificial Intelligence Profile, NIST AI 600-1 (July 2024), https://nvlpubs.nist.gov/nistpubs/ai/NIST.AI.600-1.pdf; Nat'l Inst. of Standards & Tech., Adversarial Machine Learning: A Taxonomy and Terminology of Attacks and Mitigations, NIST AI 100-2e2025 (January 2025), https://nvlpubs.nist.gov/nistpubs/ai/NIST.AI.100-2e2025.pdf.
[112] UK Biobank, *Access Procedures* (v2.1, July 2022), https://www.ukbiobank.ac.uk/media/hfnf2w0f/20220722-access-procedures-v2-1-final.pdf; UK Health Data Research Alliance, *Proposed Recommendations for a Data Use Register Standard* (Oct. 2021), https://ukhealthdata.org/alliance-outputs/.
[113] National Commission for the Protection of Human Subjects of Biomedical and Behavioral Research, *The Belmont Report: Ethical Principles and Guidelines for the Protection of Human Subjects of Research* (1979), https://www.hhs.gov/ohrp/regulations-and-policy/belmont-report/index.html.
[114] Obeid & Picone, *supra* note 67.
[115] Neural Engineering Data Consortium, *supra* note 69.

Brain foundation models strain an already difficult consent landscape.[116] Because they can be pretrained on decades of repurposed public datasets, as well as on stitched-together datasets collected for different purposes and under different protocols, they complicate the relationship between the consent terms agreed upon at the time of data collection and the ultimate downstream models and applications those data might ultimately enable.

*2.2.1. Reusing Brain Data Beyond Their Original Purpose*

A first challenge is the shift from task-specific to general-purpose models. Long before brain foundation models, consent processes for collecting and sharing data were already practically complex[117]. The use of broad consent, having participants authorize unspecified future research uses at time of data collection, meant that fairly often data subjects couldn't foresee all expected uses of their data. However, before foundation models, the ultimate uses of data were restricted by the technical reality that models required labeled datasets for training. Datasets with seizure data were necessarily channeled to models for seizure-related applications. Consequently, a for-profit med-tech company could raise questions about equity or benefit sharing by using such a dataset to build a seizure-detection system for hospitals, the reuse remained largely tethered to a foreseeable application.

This constraint has functioned as a kind of structural protection, a built-in limit on how far a model's use could stray from what was anticipated, and agreed to, at the time of a data collection. Brain foundation models effectively dissolve that constraint. Because they do not require task-specific labels and can absorb datasets collected for many different purposes, neither the task nor the ultimate application is defined when the model is trained. Indeed, that is the point of general-purpose foundation models: once pretrained, they can be adapted to a wide range of downstream uses, many of which are unknown and arguably unknowable at the time of collection. And further undermines the already tenuous benefit of broad consent.[118]

---

[116] D. J. Solove, *Privacy Self-Management and the Consent Dilemma*, 126 HARV. L. REV. 1880 (2013).
[117] N. C. Manson & O. O'NEILL, RETHINKING INFORMED CONSENT IN BIOETHICS (2007).
[118] C. Grady et al., *Broad Consent for Research with Biological Samples: Workshop Conclusions*, 15 AM. J. BIOETHICS 34 (2015)

This poses significant consent issues for the repurposing of public archives; how can brain foundation model developers ensure that individuals in public datasets collected decades ago consent to the development of these models today? For example, if a participant once consented for their de-identified EEG data to be used for research related to future research, does that mean that a research team can develop a model with the intent of using it for clinical diagnostics? This may be reasonable. But what if the developers open source it, with the goal of making it available not only for researchers and clinicians but for-profit companies developing clinical tools? And is it categorically different if a third-party surveillance firm later downloads the model and repurposes it into a workplace-monitoring product designed to track employee stress, attention, or cognitive load in real time? Neither the original participants nor the academic developers could reasonably have anticipated this trajectory, yet it is technically straightforward once a general-purpose model is made available. This also resurfaces a tension made above in relation to privacy, but now between scientific access versus the autonomy of those individuals who contributed data under particular terms.

Especially in a field oriented toward clinical benefit and innovation, there will be strong intuitive pull to treat the repurposing of legacy datasets as justified by the potential collective good as outlined in Part 1.2. Yet that intuition sits uneasily alongside the fact that many contributors never explicitly agreed to enable open-ended, general-purpose modeling, and uses for models now may conflict with the spirit of the original consent. Bioethics has seen how secondary uses can violate participant expectations even when data were collected for research.[119]

When consent cannot realistically anticipate the future, what ethical or institutional basis—if any—can legitimately authorize reuse beyond what participants expressly agreed to? This question also applies moving forward, as the field develops ex ante approaches to informed consent for brain data collection while acknowledging that ultimate downstream uses may remain unknown.

*2.2.2. Combining Datasets Collected under Different Consent Regimes*

---

[119] M. M. Mello & L. E. Wolf, *The Havasupai Indian Tribe Case—Lessons for Research Involving Stored Biologic Specimens*, 363 NEW ENGLAND J. MEDICINE 204 (2010).

A second challenge arises from stitching together datasets collected across governance regimes. Previously, training datasets typically came from a single study or institutional context, like an epilepsy clinic or a particular research consortium, ensuring a relatively coherent governance regime even when multiple sites contributed data. Today, a single model might pretrain on Institutional Review Board (IRB)−governed academic studies, HIPAA-covered clinical archives, and proprietary data from consumer neurotechnologies governed by minimal protections in terms-of-service. How comparable are these forms of consent? Should the developer be required to release the model under the strictest of the consent terms in the datasets used?

More concretely, if clinical researchers combine epilepsy recordings with consent for medical research, EEG data from consumer meditation application with broad rights for reuse, and cognitive study data for medical treatments related to cognitive decline, to train a foundation model and then, consistent with open-science norms, publicly release the model, what consent terms should actually be guiding downstream use?

*2.2.3 Agenda-Setting Questions*

1. Can consent ever be meaningfully "informed" for open-ended, future-adaptive model training and downstream repurposing? If not, what does this suggest about how consent ought to be managed?
2. How should we weigh the collective benefit of repurposing legacy datasets beyond their original, agreed-upon consent provisions, particularly given their value for medical and scientific advancement? Which stakeholders should be involved in making those determinations?
3. What governance regime should apply to models trained on datasets collected under multiple consent and governance regimes, each with its own associated rules?

*2.2.4 Baseline Safeguards*

Consent regimes for brain data were not built for foundation model training, and retrofitting or adapting them will almost certainly require multi-faceted solutions that span the full pipeline of model training and deployment. Still, certain steps are straightforward and should be implemented as a baseline. First, the field should immediately tighten consent and reuse practices for newly collected data. Consent language when collecting brain data intended for potential foundation model training should move beyond vague "future research" language and instead explicitly disclose the prospect of general-purpose pretraining, reuse across domains, and downstream fine-tuning for unknown domains, and applications that may be commercial and non-clinical. This aligns with norms in genomics for informed-consent and large-scale data-sharing.[120] For legacy datasets, developers must review the original consent terms of datasets and corresponding data use agreements, flag datasets where brain foundation model training was not contemplated, and avoid using those datasets unless additional authorization or a robust justification is in place. When datasets are stitched together from different consent regimes, model developers should treat the strictest terms as an interim default, in particular where contributors explicitly limited uses (e.g., therapeutic research only, no commercial use).[121]

## 2.3. Bias

Who is represented in training data and how models perform across different groups are central concerns in AI fairness discussions.[122] These discussions center on the ways datasets overrepresent some groups, underrepresent others, or encode assumptions about what looks normal. For brain foundation models, this matters twice over. First, they are built on brain datasets that already have well-documented demographic and structural skews. Second, by dint of being large, stitched-together, general-purpose systems, they are likely to carry their skews into many downstream uses in ways that may be hard to see or correct. Below, we lay out three interrelated issues.

### 2.3.1. Training Dataset Composition: Who Gets Measured?

---

[120] 45 C.F.R. § 46.116 (2018); Nat'l Insts. of Health, *NIH Genomic Data Sharing Policy*, 79 Fed. Reg. 51345 (August 28, 2014), https://www.federalregister.gov/documents/2014/08/28/2014-20385/final-nih-genomic-data-sharing-policy.

[121] J. Lawson et al., *The Data Use Ontology to Streamline Responsible Access to Human Biomedical Datasets*, 1 CELL GENOMICS 100028 (2021); J. Kaye et al., *Dynamic Consent: A Patient Interface for Twenty-First Century Research Networks*, 23 EUR. J. HUM. GENETICS 141 (2015).

[122] S. Barocas & A. D. Selbst, *Big Data's Disparate Impact*, 104 CAL. L. REV. 671 (2016); J. Buolamwini & T. Gebru, *Gender Shades: Intersectional Accuracy Disparities in Commercial Gender Classification*, in PROCEEDINGS OF THE 1ST CONFERENCE ON FAIRNESS, ACCOUNTABILITY AND TRANSPARENCY 77 (2018); T. Bolukbasi et al., *Man Is to Computer Programmer as Woman Is to Homemaker? Debiasing Word Embeddings*, in ADVANCES IN NEURAL INFORMATION PROCESSING SYSTEMS 29 (2016).

Long before brain foundation models, neuroethics research showed that brain datasets are not neutral snapshots of the generic human brain but products of who gets measured. Clinical datasets drawn from hospitals overrepresent affluent, insured, and, often, White patients.[123] Academic EEG and fMRI studies skew toward Western, educated, industrialized, rich, and democratic (WEIRD) populations.[124] Consumer neurotechnology data tends to come from people who can afford the devices and are willing to wear them.[125] Many of the datasets that will feed brain foundation models are therefore already skewed.

Brain foundation models will be trained by repurposing existing datasets, such as those from public research repositories, legacy clinical archives, and open benchmarks, alongside new data streams from consumer hardware and platforms. If developers simply use readily available datasets without asking who those datasets represent, they are likely to reproduce the same gaps at a much larger scale. Stitching datasets together does not automatically solve the problem, as each set brings its own recruitment practices and barriers to participation. Larger sizes may just layer skewed samples, and whether stitching improves representation is an empirical question.[126]

This creates a concrete challenge for brain foundation model developers. They must make the composition of large, stitched training corpora visible enough to scrutinize, and they must develop tools for assessing who is actually represented. For legacy public datasets, this may require working with incomplete or inconsistent metadata. An analysis of 11.4 million sample records in two major biomedical repositories found that most metadata field names and values are neither standardized nor controlled, and nearly all attribute names were custom or ad hoc rather than drawn from official dictionaries.[127] For new, especially commercial datasets, it raises questions about how sampling decisions are made and documented in the first place.

---

[123] L. Rutten-Jacobs et al., *Racial and Ethnic Diversity in Global Neuroscience Clinical Trials*, 4 CONTEMPORARY CLINICAL TRIALS COMMUNICATIONS 101255 (2024).
[124] T. Choy et al., *Systemic Racism in EEG Research: Considerations and Potential Solutions*, 26 AFFECTIVE SCIENCE 14 (2021); Joseph Henrich et al., *The Weirdest People in the World?*, 33 BEHAVIORAL & BRAIN SCIENCES 61 (2010).
[125] A. Wexler, *Who Uses Direct-to-Consumer Brain Stimulation Products, and Why? A Study of Home Users of tDCS Devices*, 2 J. COGNITIVE ENHANCEMENT 114 (2018).
[126] A. Khuntia et al., *Sampling Inequalities Affect Generalization of Neuroimaging-Based Diagnostic Classifiers in Psychiatry*, 21 BMC MED. 241 (2023)
[127] R. S. Gonçalves & M. A. Musen, *The Variable Quality of Metadata About Biological Samples Used in Biomedical Experiments*, 6 SCI. DATA 190021 (2019)

*2.3.2. Model Performance: Hardware, Accuracy, and Neurotypical Norms*

At the model level, we already see performance gaps in task-specific EEG models. Supervised classifiers trained for tasks like Parkinson's detection have shown markedly different results across groups, with accuracy falling from 81% for men to 64% for women.[128] In these cases, it is hard to fully disentangle whether the gap comes from recruitment, labeling, architecture, or something else, but the effect is clear: some groups get systematically better predictions and tools than others. On the hardware side, EEG caps and sensors tend to perform better on some people's bodies than others. Many current systems do a poor job of accommodating hairstyles, scalp colors, or ear shapes, leading to longer setup times, poor signal quality, or exclusion from studies altogether.[129] Even when people are technically in the dataset, their data may be noisier, sparser, or missing in ways that will matter for training and evaluation.

Because of their nature, brain foundation models will be trained on exactly this mix of uneven signals and performance. On one hand, their ability to learn from vast quantities of messy, in-the-wild EEG and other signals may help them handle lower-quality data better than today's narrow models, including consumer neurotech devices. On the other hand, foundation models trained on biased inputs may not wash those biases out. In some cases, such as multimodal models that may include different types of brain data, combining modalities can sometimes compound and amplify biases already present in the underlying datasets.[130]

Finally, these systems also risk standardizing a particular profile of attention, affect, or cognition as "normal."[131] When models trained on narrow samples are used to support clinical decisions, aid in classrooms, or monitor workplaces, they can treat the learned baseline as the standard of neurological health or productivity. People whose neural patterns diverge are systematically misread as inattentive, noncompliant, or impaired. In the neuro context, this could manifest as a model highly accurate at detecting Alzheimer's in well-represented Western patient groups and markedly less reliable for patients from underrepresented populations.

---

[128] A. Kurbatskaya et al., *Assessing Gender Fairness in EEG-Based Machine Learning Detection of Parkinson's Disease: A Multi-Center Study*, in 2023 31ST EUROPEAN SIGNAL PROCESSING CONFERENCE (EUSIPCO) (2023).
[129] Choy et al., *supra* note 124.
[130] M. Drissi, *More Is Less? A Simulation-Based Approach to Dynamic Interactions Between Biases in Multimodal Models*, arXiv (2024), https://arxiv.org/abs/2412.17505.
[131] "RAW DATA" IS AN OXYMORON (L. Gitelman ed., 2013); G.C. Bowker & S.L. Star, *Sorting Things Out: Classification and Its Consequences* (2000).

Therefore, whether brain foundation models narrow or widen performance gaps across demographics, hardware types, and neurotypes is an empirical question. We have almost no systematic bias evaluations for brain foundation models yet, but experience with LLMs suggests we should assume skew.[132]

*2.3.3. General-Purpose Models as Bias Multipliers*

Finally, brain foundation models raise the fairness risk of downstream amplification. For task-specific models, if a single training dataset is skewed, the harm is largely contained to that one model. However, a brain foundation model trained on a particular mix of data can have those same biases amplified and propagated as it is fine-tuned and reused across many settings. The same model, developed with a specific training dataset and its gaps, can end up underpinning workplace monitoring tools, educational assessment systems, and clinical decision support.[133]

In that sense, the worry is not just that bias exists at the training stage but that it travels, skewing who is represented or how "normal" brain function is encoded, which can quietly shape decisions in domains far removed from the original context of data collection.

*2.3.4 Agenda-Setting Questions*

1. How should training-data representativeness be defined and operationalized for brain foundation models, such that datasets and models do not implicitly encode a single "normal" or neurotypical brain as the default? What techniques are needed to make diversity in brain data legible, measurable, and auditable?
2. How can we disentangle whether performance gaps stem from hardware, data quality, data labeling, or model architecture?

---

[132] E. M. Bender et al., *On the Dangers of Stochastic Parrots: Can Language Models Be Too Big?*, in PROCEEDINGS OF THE 2021 ACM CONFERENCE ON FAIRNESS, ACCOUNTABILITY, AND TRANSPARENCY 610 (2021).
[133] *See* Bommasani et al., *supra* note 2; Bender et al., *supra* note 132.

3. What auditing and evaluation mechanisms are needed to detect bias in brain foundation models, especially as they are trained as multimodal models? What documentation should be required to travel from training data through release and deployment to make that bias visible and accountable?

*2.3.5 Baseline Safeguards*

Addressing bias systematically will require both technical and socio-technical research, focused on the whole "stack" , from developing stable representations of human and neural variability, to the hardware layer, dataset layer, and the real-time model deployment contexts. However, there are established methods to mitigate bias within the AI field that can be implemented for brain foundation models immediately. First, dataset and model developers should create robust and provenance documentation in standardized formats, such as datasheets,[134] detailing dataset demographic composition, context of collection, and device metadata, as well as what is systematically absent. Developers should also evaluate model performance across relevant subgroups and conditions, not just aggregate benchmarks[135], with the recognition that today's subgroup categories are provisional and will evolve over time. Evaluation should explicitly distinguish, to the extent possible, whether disparities arise from sensor performance, recruitment and consent patterns, labeling practices, or modeling choices and should report where disentanglement is not currently possible.[136] Developers should also evaluate *extrinsic* bias, i.e., bias that emerges in downstream use after fine-tuning, transfer, and deployment, rather than treating pretraining benchmarks as sufficient.[137] Finally, when documentation is insufficient or when evaluation results indicate systematic performance gaps, deployment should be limited until they are resolved.[138]

---

[134] T. Gebru et al., *Datasheets for Datasets*, 64 COMMC'NS ACM 86 (2021)
[135] M. Mitchell et al., *Model Cards for Model Reporting*, in PROCEEDINGS OF THE 2019 ACM CONFERENCE ON FAIRNESS, ACCOUNTABILITY, AND TRANSPARENCY 220 (2019); U.S. Food & Drug Admin., Health Canada & U.K. Medicines & Healthcare Prods. Regulatory Agency, *Good Machine Learning Practice for Medical Device Development: Guiding Principles* (Oct. 2021).
[136] S. Barocas et al., *Designing Disaggregated Evaluations of AI Systems: Choices, Considerations, and Tradeoffs*, in PROCEEDINGS OF THE 2021 AAAI/ACM CONFERENCE ON AI, ETHICS, AND SOCIETY 368 (2021)
[137] Y. Guo et al., *Bias in Large Language Models: Origin, Evaluation, and Mitigation*, arXiv (2024), https://arxiv.org/abs/2411.10915.
[138] NIST AI 100-1, *supra* note 111.

**2.4. Benefit Sharing and Equitable Data Access**

One person's brain patterns cannot single-handedly underpin an AI system. However, when datasets composed of tens of thousands of individuals' neural recordings are aggregated across medical, research, and consumer contexts, they can collectively underwrite highly valuable commercial systems. The questions, then, are who gets to build on that collective resource and who shares in the benefits.

In the recent discourse following the rollout of LLMs, creators and journalists were furious that companies had effectively vacuumed up decades of expressive work to train proprietary models. Fairness was mostly cast in economic terms, such as copyright, licensing, and compensation for creative labor.[139] Brain foundation models are now raising similar questions. Here, however, the raw material is not only expressive output but also bodies and health, sitting squarely in a set of problems that bioethics has been wrestling with for decades in the context of tissue samples, genetic information, and biobanks.[140]

In what follows, we focus on public versus private benefit sharing, competing conceptions of fairness in benefit sharing, and equitable data access for model developers.

*2.4.1. Public versus Private Benefit Sharing*

Many of the brain datasets likely to feed brain foundation models have some public or collective backing, such as research grants, national health systems and teaching hospitals, or philanthropic support.[141] Ethics documents, grant proposals, and recruitment materials typically frame participation as contributing to science or improving care, not as supplying inputs for future proprietary AI models.

---

[139] K. Lee et al., *Talkin' 'Bout AI Generation: Copyright and the Generative-AI Supply Chain*, arXiv (2023), https://arxiv.org/abs/2309.08133.
[140] Human Genome Organisation Ethics Comm., *Statement on Benefit Sharing* (Apr. 9, 2000), *reprinted in* 3 COMMUNITY GENETICS 88 (2000); K. S. Steinsbekk et al., *We're Not in It for the Money—Lay People's Moral Intuitions on Commercial Use of 'Their' Biobank*, 16 MEDICINE, HEALTH CARE & PHILOSOPHY 151 (2013).
[141] Human Genome Organisation Ethics Comm., *supra* note 140; K. S. Steinsbekk & B. Solberg, *Biobanks—When Is Re-Consent Necessary?*, 4 PUB. HEALTH ETHICS 236 (2011); P. Tindana et al., *'It Is an Entrustment': Broad Consent for Genomic Research and Biobanks in Sub-Saharan Africa*, 19 DEVELOPING WORLD BIOETHICS 9 (2019).

The world has seen tech companies' research labs leverage public datasets as part of their research. For example, Meta's 2023 brain-decoding work drew not only on internal resources but also on THINGS-MEG, an open academic magnetoencephalography (MEG) dataset built through years of publicly oriented research.[142] That kind of academic–industry pipeline is not inherently problematic; there are certain benefits to these collaborations, like that commercializing technology makes it accessible to broad populations.

The question is how to structure the arrangement fairly. With text- and image-based models, public outcry over generative AI has already pushed companies toward a patchwork of responses, including private licensing deals, opt-out mechanisms, and experiments with synthetic training data. Brain foundation models now pose a similar challenge. When public and academic brain datasets become valuable inputs to proprietary models, what is owed to the institutions and communities that produced them? Should there be concrete forms of return, like access, remuneration, or shared infrastructure?[143] Moreover, what would a fair arrangement look like when models are built on collective neural resources?

*2.4.2. Economic versus Dignitarian Lens on Benefit Sharing*

What would fair compensation or benefit sharing look like for people whose brain data help train brain foundation models? As mentioned, in the LLM context, this question has been framed in economic—and largely intellectual property—terms, and it is tempting to import that logic. If brain foundation models generate commercial value, should some of that value not flow to the contributors or institutions that supplied the data?

This could involve returns to the hospitals and universities that built the datasets, preferential access for contributing health systems, or even individual-level payments when a person's recordings enter a training corpus. However, a purely economic frame sits uneasily for brain data. Unlike text or images, these datasets are collected in contexts structured by clinical and research norms rather than market exchanges. Debates in biobanking and genomics have already shown how ill-fitting it is to treat bodily materials as owned commodities.[144] Simply paying individuals for their brain data risks treating bodies and health histories as income-generating assets, and it may place particular pressure on marginalized people to sell something they would otherwise have kept private.[145] And while legal precedent like *Moore v. Regents* have denied patients property rights over their tissue, even when it underpins lucrative products[146], this does not settle the normative question. It simply cannot be the case that contributors have no claim whatsoever when their brain data, arguably an extension of their bodily identity, generates profit for companies.

---

[142] Y. Benchetrit et al., *Brain Decoding: Toward Real-Time Reconstruction of Visual Perception*, in INT'L CONF. ON LEARNING REPRESENTATIONS (ICLR) (2024); M. N. Hebart et al., *THINGS-Data, a Multimodal Collection of Large-Scale Datasets for Investigating Object Representations in Human Brain and Behavior*, 12 ELIFE e82580 (2023).
[143] Steinsbekk & Solberg, *supra* note 141; D. Hallinan, *Biobank Oversight and Sanctions Under the General Data Protection Regulation*, in GDPR AND BIOBANKING: INDIVIDUAL RIGHTS, PUBLIC INTEREST AND RESEARCH REGULATION ACROSS EUROPE 121 (2021); B. Prainsack & A. Buyx, *A Solidarity-Based Approach to the Governance of Research Biobanks*, 21 MED. L. REV. 71 (2013).

For brain foundation models, then, the stakes are both economic and dignitarian. On the economic side, there are questions about who captures the value produced from publicly backed brain datasets and whether some of that value should be reinvested in the institutions and communities that produced them. On the dignitarian side, there are questions about respect, bodily integrity, collective self-determination.[147] For example, who gets a say in how shared neural resources are used and on what terms? This is why work on biobanks and health data governance has increasingly turned to more structural, collective stewardship arrangements,[148] including data trusts[149], data intermediaries,[150] and data solidarity models.[151]

## 2.4.3 Procedural Equity in Data Access

We have discussed equitable distribution of benefit for data subjects whose data is used to train the models, or the public who is funding the development of the datasets. In this section, we turn attention to a different stakeholder; the model developers. In particular, we consider what conditions must be in place in order to ensure equitable access to data for the different individuals and institutions that want access to public datasets.

---

[144] D. Satz, WHY SOME THINGS SHOULD NOT BE FOR SALE: THE MORAL LIMITS OF MARKETS (2010).
[145] Satz, *supra* note 144.
[146] Moore v. Regents of Univ. of Cal., 793 P.2d 479 (Cal. 1990).
[147] E. Seger et al., *Democratising AI: Multiple Meanings, Goals, and Methods*, in 2023 AAAI/ACM CONFERENCE ON AI, ETHICS, AND SOCIETY 715 (2023)
[148] Ada Lovelace Inst., *Participatory Data Stewardship: A Framework for Involving People in the Use of Data* (Sept. 7, 2021), https://www.adalovelaceinstitute.org/report/participatory-data-stewardship/.
[149] Ada Lovelace Inst. & AI Council, *Exploring Legal Mechanisms for Data Stewardship* (Mar. 2021); S. Delacroix & N. D. Lawrence, *Bottom-up Data Trusts: Disturbing the 'One Size Fits All' Approach to Data Governance*, 9 INT'L DATA PRIVACY L. 236 (2019).
[150] A. Giannopoulou et al., *Intermediating Data Rights Exercises: The Role of Legal Mandates*, 12 INT'L DATA PRIVACY L. 316 (2022).
[151] Prainsack & Buyx, *supra* note 143.

Open science depends on knowledge exchange and data sharing.[152] Access to large datasets is critical for the research community to support reproducibility and method benchmarking and evaluation, as well as for repurposing and reuse for further studies.[153] Easy access to large datasets is often cited as a significant driver of medical and scientific innovation. Consider how the Royal Swedish Academy credited the Protein Data Bank as instrumental to the research around computational protein structure that won the 2024 Nobel Prize in Chemistry.[154] But open data advocates point to the risks associated with the project, especially when it comes to sensitive medical data, and genomic or neural data. As described earlier in the privacy and consent sections, the availability of weakly deidentified data amplifies the risks of linking people to sensitive medical information or use of data beyond original consent terms.

In practice, many sensitive neuroscience and health datasets are not fully open; access is often conditional on credentialing, applications, and data-use agreements, often administered by repositories or data access committees.[155] These controls can reduce misuse risk, but they can also introduce procedural friction (applications, unclear eligibility criteria, opaque review), that delays access and can systematically favor institutions with dedicated administrative/legal capacity or generally well-resourced actors.[156] If there is ambiguity around what constitutes 'qualified researchers,' it can advantage better-resourced applicants who can navigate complex access arrangements.

---

[152] S. Choudhury et al., *Big Data, Open Science and the Brain: Lessons Learned from Genomics*, 8 FRONTIERS IN HUMAN NEUROSCIENCE 239 (2014).
[153] Markiewicz et al., *supra* note 69.
[154] Nobel Comm. for Chemistry, *Scientific Background to the Nobel Prize in Chemistry 2024: Computational Protein Design and Protein Structure Prediction*, at 11 (Oct. 9, 2024), https://www.nobelprize.org/uploads/2024/10/advanced-chemistryprize2024.pdf.
[155] See, e.g., Markiewicz et al., *supra* note 69; C. Sudlow et al., *UK Biobank: An Open Access Resource for Identifying the Causes of a Wide Range of Complex Diseases of Middle and Old Age*, 12 PLOS MEDICINE e1001779 (2015); Petersen et al., *supra* note 77.
[156] M. Shabani et al., *From the Principles of Genomic Data Sharing to the Practices of Data Access Committees*, 7 EMBO MOLECULAR MEDICINE 507 (2015); F. McKay et al., *Artificial Intelligence and Medical Research Databases: Ethical Review by Data Access Committees*, 24 BMC MEDICAL ETHICS 1 (2023).

*2.4.4 Agenda-Setting Questions*

1. What benefit-sharing expectations should we hold commercial actors to when they train models on publicly funded brain datasets?
2. If an economic model of remuneration, such as individual compensation, is the wrong model for brain data, what collective governance mechanisms are appropriate? What institutions are best positioned to weigh in on and set those terms?
3. Should benefit-sharing obligations attach to models (not just datasets), and if so, how would that be operationalized as models are fine-tuned, redistributed, and deployed downstream?
4. How should we balance the benefits of open access to data with the potential costs, delays, or friction introduced by necessary governance and restrictions?

*2.4.5 Baseline Safeguards*

Some of the questions that we have raised are novel and will require sustained multi-stakeholder deliberation, however, certain defaults can be put in place now both to establish a floor and to avoid the public and policy backlash that followed the emergence of LLMs. First, dataset developers should adopt tiered access defaults for sharing (e.g., tiered eligibility, public-interest review, secure environments) rather than fully open or fully closed access.[157] Second, model developers should specify in concrete and auditable terms, what forms of return flow to the communities and institutions that generated, stewarded, and curated the underlying data. Those returns could take multiple forms, such as academic access tiers, subsidized pricing in under-resourced settings, or shared infrastructure.[158] Third, for-profit neurotechnology companies, open neuroscience groups, and clinical and academic labs, should prioritize coordinated infrastructure and stewardship arrangements that prevent individual payment from becoming the default model for brain data, given the dignitarian stakes of its commodification. This can initially involve aligning with (or extending) existing data access committees, data trusts, and other data intermediaries.

---

[157] UK Biobank, *supra* note 112.
[158] Council for Int'l Orgs. of Med. Scis. & World Health Org., I*nternational Ethical Guidelines for Health-Related Research Involving Humans* (4th ed. 2016); A. Bedeker et al., *A Framework for the Promotion of Ethical Benefit Sharing in Health Research*, 7 BMJ GLOBAL HEALTH e008096 (2022).

## 2.5. An Unsettled Governance Landscape

Before considering how brain foundation models *should* be governed, it is worth asking what governance regime they currently fall under, if any. The answer is not obvious. Like many emerging technologies, brain foundation models do not fit neatly into existing legal frameworks, which were designed for earlier technical realities.

The picture is further complicated by the fact that foundation models and neural data are each already subjects of unresolved legal debate. Foundation models raise questions about copyright, training data ethics, and deployment. Neural data has become a focus of state, federal, and international regulatory efforts. Brain foundation models sit at the intersection, inheriting the unsettled questions of both.

We begin by first addressing the absence of a unified legal framework for brain foundation model governance, then turn to the current patchwork of frameworks that do appear to govern this space, albeit insufficiently.

### 2.5.1. The Missing Copyright Analog

In the years after generative AI companies launched their models, we witnessed a flurry of lawsuits from the aggrieved parties mentioned earlier. Three well-known authors filed class action suits against OpenAI and Meta, alleging that their books were copied into the training data without permission.[159] *The New York Times* sued OpenAI and Microsoft for the unpermitted use of its articles to train GPT models.[160] Getty Images sued Stability AI, alleging that the company had used millions of Getty images to train Stable Diffusion without purchasing a license.[161] In each of these cases, the plaintiffs invoked copyright law, arguing that the technology companies had harvested terabytes of copyrighted text and images to pretrain their text and image foundation models.[162] While there are differing opinions as to whether copyright ought to govern

---

[159] S. Levin, *Sarah Silverman Sues OpenAI and Meta for Copyright Infringement*, THE GUARDIAN (July 10, 2023), https://www.theguardian.com/technology/2023/jul/10/sarah-silverman-sues-openai-meta-copyright-infringement.
[160] Getty Images (US), Inc. v. Stability AI, Ltd., No. 1:23-cv-00135 (D. Del. filed Feb. 6, 2023).
[161] K. Chan & M. O'Brien, *Getty Images and Stability AI Face Off in British Copyright Trial That Will Test AI Industry*, ASSOCIATED PRESS (June 9, 2025), https://apnews.com/article/getty-images-stability-ai-copyright-trial.
[162] *See* Silverman v. OpenAI, No. 3:23-cv-03417, ECF No. 1, at 2–4 (N.D. Cal. July 7, 2023); Getty Images (US), Inc. v. Stability AI, Ltd., No. 1:23-cv-00135 (D. Del. filed Feb. 6, 2023); The New York Times Co. v. OpenAI, Inc.,

these works if used as training datasets for AI,[163] it nonetheless gave creators a clear, familiar doctrinal hook to pull on.

There is no obvious single legal mechanism for brain data as training data. Brain recordings are not expressive works as would be covered by copyright, and there is no precedent for individual ownership rights over brain data or biological samples, as *Moore v. Regents of University of California* made clear in the case of discarded tissue.[164] In short, even though there is repurposing of people's public data to train foundation models, there is no single uniform protection we can turn to. This raises the question: if not copyright, what actually governs this space, and are those existing tools collectively sufficient?

## 2.5.2. The Current Regulatory Patchwork and Future Governance Paths

Brain foundation models do not occupy a legal vacuum, they sit within a patchwork of existing frameworks, each with limitations when applied to training data governance.

First, there is sectoral oversight: health privacy laws like HIPAA and research ethics frameworks like IRBs. These regimes tie protections to the physical or institutional context in which data is collected. But the protections do not follow the data; once it moves to a new context, the original governance regime no longer applies.[165] As discussed in section 2.1.1, clinical and research datasets are often released publicly after de-identification. Once released, the original governance regime, whether HIPAA or IRB, no longer applies.

Second, a new wave of neural data-specific protections has emerged,[166] including amendments to consumer privacy laws in several U.S. states and proposed federal legislation like the proposed

---

No. 1:23-cv-11195 (S.D.N.Y. filed Dec. 27, 2023); *see also* Michael M. Grynbaum & Ryan Mac, *The Times Sues OpenAI and Microsoft Over A.I. Use of Copyrighted Work*, N.Y. TIMES (Dec. 27, 2023).
[163] M.A. Lemley & B. Casey, *Fair Learning*, 99 TEX. L. REV. 743 (2021); Lee et al., *supra* note 139.
[164] Moore v. Regents of Univ. of Cal., *supra* note 146.
[165] Nissenbaum, *supra* note 110; S. Barocas & H. Nissenbaum, *Big Data's End Run Around Procedural Privacy Protections*, 57 COMMC'NS ACM 31 (2014)
[166] This also includes international efforts like constitutional neurorights provisions in Chile and elsewhere, international frameworks like UNESCO's recommendation on neurotechnology. See Constitución Política de la República de Chile [C.P.] art. 19, no. 1 (as amended by Ley No. 21.383, Oct. 25, 2021) (Chile); United Nations Educ., Sci. & Cultural Org., *Recommendation on the Ethics of Neurotechnology*, UNESCO Doc. SHS/BIO/REC-NEURO/2025 (Nov. 2025)

Management of Individuals' Neural Data (MIND) Act.[167] These laws treat neural data as a protected category of 'sensitive information', and govern data collected from commercial wearables, one source of proprietary training data for brain foundation models.[168] But what these laws actually protect remains uncertain.[169] It is not evident, for instance, whether each one would cover inferences drawn from neural data or only raw signals, and how that distinction would bear on data used in model training. More fundamentally, these laws vary but are for the most part limited in scope. They apply to proprietary data collected in consumer settings, not to open science datasets, clinical archives, or data collected by invasive BCI companies.[170] Much of the training data ecosystem, in other words, falls outside their reach.

Third, there are emerging risk-based approaches to AI governance, such as the EU AI Act.[171] The Act classifies certain AI systems as high-risk based on their application and imposes corresponding transparency and safety requirements.[172] Brain foundation models used in healthcare or employment could, in principle, fall under this designation. However, the Act focuses on deployment, not upstream training. Additionally, it does not address how to govern models that develop their capabilities through cross-context training on datasets stitched together from clinical, research, and consumer sources.

If these frameworks, taken together, addressed the concerns raised in this paper, the absence of a unified legal framework for brain data might not matter. Whether they do, however remains unclear.

Proponents of regulatory patchworks, point to their value as 'laboratories of democracy', environments in which jurisdictions experiment with different rules and drive innovation through competition,[173] but that is not the patchwork we are describing. The fragmentation here runs across jurisdictions, sectors, technology types, and stages of the AI pipeline, and at first glance appears to represent less a productive experiment than a series of gaps. This might be unavoidable, given the unprecedented capabilities of brain foundation models and other general-purpose AI systems. And in fact, scholars have recently suggested that these complex dynamics may require a new approach to governance entirely, one that moves beyond siloed sectoral laws and anticipates the categorical instability of converging data technologies and contexts.[174]

---

[167] Colo. H.B. 24-1058, 74th Gen. Assemb., 2d Reg. Sess. (2024) (amending Colo. Rev. Stat. § 6-1-1303); Cal. S.B. 1223, 2023–2024 Leg., Reg. Sess. (Cal. 2024) (amending Cal. Civ. Code § 1798.140); Mont. S.B. 163, 69th Leg., Reg. Sess. (2025) (amending Mont. Code Ann. § 10-13-804); Conn. S.B. 1295, Gen. Assemb., Jan. Sess. (2025) (amending Conn. Gen. Stat. § 42-515 et seq.).
[168] The MIND Act also covers "other related data", a class of data defined to include biometric, physiological, or behavioral information that, while not directly measuring neural activity, can be processed or combined with other data to infer cognitive, emotional, or psychological states. See MIND Act of 2025, S. 2925, 119th Cong. § 2(5) (2025).
[169] N. Gupta, *What Are Neural Data? An Invitation to Flexible Regulatory Implementation*, STAN. L. SCH. L. & BIOSCIENCES BLOG (Dec. 2, 2024); Magee et al., *supra* note 94.
[170] Colo. Rev. Stat. § 6-1-1304(2)(a), (d), (o) (2024); Cal. Civ. Code § 1798.145(c)(1)(A), (j) (West 2024).
[171] Regulation (EU) 2024/1689 of the European Parliament and of the Council of 13 June 2024 Laying Down Harmonised Rules on Artificial Intelligence (Artificial Intelligence Act), 2024 O.J. (L 1689) 1.
[172] Regulation (EU) 2024/1689, arts. 6–7 & Annex III, 2024 O.J. (L 1689) 1.

*2.5.3 Agenda-Setting Questions*

1. What is the appropriate target of regulation for brain foundation models: data type, training practices, model release, inferences enabled, or downstream uses?
2. Where should governance obligations attach across the pipeline of model development? On the dataset developers, repositories, model developers, or model deployers? Where within this pipeline is there sufficient governance and where requires more?
3. What minimum transparency/provenance requirements are needed to make oversight possible?
4. What institutional forums should convene and maintain standards across neuroscience, ethics, disability communities, industry, and policy?

*2.5.4 Baseline Safeguards*

While the policy and legal questions raised in this section will require significant attention by the field and may even require new cross-domain governance frameworks,[175] baseline governance practices, beyond those outlined in the preceding sections, can and should be established now. First, dataset and model developers should adopt participatory design methods that bring affected communities into scoping, design, and evaluation of brain foundation model development. It is critical that this happens early and often, not after the fact.[176] Second, developers should adopt an explicit, reviewable rationale for model release type, with input from policymakers and researchers. Model release decisions function as a form of governance and whether models are released with open weights, API, or controlled access has implications for appropriate training dataset governance.[177] Third, at the field level, durable review forums should be established, spanning neuroscience, clinical practice, disability and neurodivergent communities, industry, and policy, to support ongoing standard-setting that can evolve alongside empirical evidence, downstream deployments, and policy and lawmaking.[178] Finally, to make that adaptive governance possible, researchers should produce translational materials, such as white papers, technical briefs, and policy explainers that synthesize technical and operational realities for review forums and policymakers. These materials should track capabilities, developers, active deployments, and concrete failure modes and harms.[179]

---

[173] New State Ice Co. v. Liebmann, 285 U.S. 262, 311 (1932) (Brandeis, J., dissenting); M. C. Dorf & C. F. Sabel, *A Constitution of Democratic Experimentalism*, 98 COLUM. L. REV. 267 (1998).
[174] N. Farahany, *2026: The Year Everything Converges*, THINKING FREELY (Dec. 2025), https://nitafarahany.substack.com/p/2026-the-year-everything-converges.
[175] Farahany, *supra* note 174.

**CONCLUSION**

Brain foundation models are just beginning to enter broader awareness. As we await the downstream application of this nascent technology (with cautious optimism about their ability to help relieve longstanding challenges in medical diagnosis and treatment or neuroscience research), we focus this paper on a set of ethical and governance questions already unfolding at the training data stage. In particular, in this paper, we demonstrate how the intersection of the foundation model paradigm with neural data, or brain foundation models, result in a normatively distinctive upstream training-data environment, and one that has received little systematic ethical analysis.

In Part 1, we provided a descriptive account of brain foundation models, including how they work and how they differ from supervised, task-specific models, and we outlined the downstream settings in which they are beginning to be developed and piloted. We also characterized the training-data ecosystem underpinning the field; i) the public and proprietary datasets that support model development, ii) the actors who generate and steward those data, and iii) the concrete practices through which developers are repurposing public datasets and combining them with multimodal and proprietary datasets. In Part 2, we used that technical and ecosystem grounding to map the ethical, social, and legal concerns raised at the training-data layer, organizing the landscape across privacy, consent, bias, benefit sharing/equitable access, and governance. Given the scope and complexity of these issues, we did not attempt to answer or solve them; rather, within each section we surfaced critical questions for the field to engage with and offered a baseline set of "raise-the-floor" practices intended to guide future empirical work, technical evaluation, and governance design as these systems mature.

---

[176] D. Saxena et al., *Emerging Practices in Participatory AI Design in Public Sector Innovation*, in PROCEEDINGS OF THE EXTENDED ABSTRACTS OF THE CHI CONFERENCE ON HUMAN FACTORS IN COMPUTING SYSTEMS 777:1 (2025).
[177] I. Solaiman et al., *Release Strategies and the Social Impacts of Language Models*, arXiv (2019), https://arxiv.org/abs/1908.09203; I. Solaiman, *The Gradient of Generative AI Release: Methods and Considerations*, in 2023 ACM CONFERENCE ON FAIRNESS, ACCOUNTABILITY, AND TRANSPARENCY 111 (2023).
[178] UKHDRA, *supra* note 112; UK Biobank, *supra* note 112.
[179] Org. for Econ. Coop. & Dev., *Framework for Anticipatory Governance of Emerging Technologies*, OECD Sci., Tech. & Indus. Pol'y Papers No. 165 (Apr. 2024)

Many of the problems mapped in this paper are not new. Risks posed to individuals when weakly de-identified datasets are made available to third parties are longstanding in the literature; consent has been critiqued for decades across bioethics, neuroethics, and data-driven systems as an insufficient mechanism for protecting individuals or ensuring meaningful autonomy. Nor is it new that private benefit can be built on publicly generated resources, or that emerging technologies do not fit neatly into existing regulatory frameworks. Brain foundation models bring many of these concerns together in a way that makes them especially acute and difficult to address.

There is some urgency to considering these questions now. Even before brain foundation models become prevalent, and before we confront downstream harms, they may reshape upstream incentives: encouraging the collection of neural data at scale, or the creation of markets for buying and selling it. Unlike before, the data need not be standardized or labeled to be useful to a range of actors, and it can consequently be aggregated, traded, and repurposed with far less friction than ever before. Companies are already beginning to market 'brain data as a service,' offering a glimpse of a future in which neural data is productized at scale.

Our concern is not that data sharing is inherently wrong, or that all forms of commercial exchange should be categorically unacceptable. It is that these practices should not proceed without meaningful scrutiny, public deliberation, and certainly not as an unexamined, default trajectory. For-profit neurotechnology companies will increasingly find that their data, collected from consumer wearables and invasive BCI alike, is its own asset and potential independent revenue stream. Those incentives intensify further if neural data (or models trained on it) become a standard input for improving general AI systems via fine-tuning, evaluation, or multimodal integration.

Many of the hardest questions outlined in this paper will require sustained research and dialogue across research, industry, and policy. We offer these as guides for a longer and broader research agenda for the field. And we proposed a "floor" of practices, which can be put in place now while the field navigates and debates the larger questions. This floor should be treated as a shared responsibility across dataset designers who still have control of their datasets, repositories and archives that host datasets, model developers across research and industry, clinical and research institutions, and commercial neurotechnology firms, in dialogue with policymakers and affected communities.

Brain foundation models constitute a promising and rapidly developing field, but if brain foundation models are to mature responsibly, and if public trust in underlying research and clinical infrastructures, not to mention newly emerging neurotechnology systems, is to be sustained, then upstream training-data governance has to become a first-order design and policy concern, rather than an afterthought.